\documentclass[aps,pra,reprint]{revtex4-1}

\usepackage{amsmath,amssymb,graphicx}
\usepackage{array}
\usepackage[caption=false]{subfig}
\usepackage{bm}
\usepackage{amsfonts}
\usepackage{epsfig,epstopdf}
\usepackage{subfig}
\usepackage{url}
\usepackage{mathtools}

\def\Tr{{\rm Tr}}

\def\la{\left\langle}
\def\ra{\right\rangle}

\def\br{{\bf r}}

\def\bA{{\bf A}}
\def\bB{{\bf B}}

\def\bF{{\bf F}}

\def\bu{{\bf u}}

\def\hrho{\hat \rho}

\def\bH{{\bf H}}

\def\cH{{\cal H}}
\def\bS{{\bf \Sigma}}

\def\be{\begin{equation}}
\def\ee{\end{equation}}
\def\ba{\begin{align}}
\def\nn{\nonumber\\}

\def\defeq{\buildrel \rm def \over =}

\def\bs{{\bf s}}
\def\b0{{\bf 0}}
\def\cA{{\cal A}}
\def\cL{{\cal L}}
\def\la{\left\langle}
\def\ra{\right\rangle}

\def\br{{\bf r}}
\def\bs{{\bf s}}

\def\bu{{\bf u}}

\def\bA{{\bf A}}
\def\bB{{\bf B}}

\def\bF{{\bf F}}

\def\bH{{\bf H}}

\def\bl{\bm{l}}

\def\cH{{\cal H}}

\def\bS{{\bf \Sigma}}

\def\Tr{{\rm Tr\ }}

\def\be{\begin{equation}}
\def\ee{\end{equation}}
\def\ba{\begin{align}}
\def\nn{\nonumber\\}
\def\la{\langle}
\def\ra{\rangle}

\def\defeq{\buildrel \rm def \over =}

\def\cR{{\cal R}}

\def\cD{{\cal D}}

\def\Cl{C_\lambda}

\def\bs{{\bf s}}
\def\br{{\bf r}}
\def\brp{{\bf r}'}
\def\bA{{\bf A}}
\def\bB{{\bf B}}
\def\xpp{x^{\prime\prime}}
\def\pmu{\partial_\mu}
\def\pnu{\partial_\nu}
\def\bsigma{\bm{\sigma}}

\begin{document}

\title{Quantum Limited Superresolution of Extended Sources in One and Two Dimensions}

\author{Sudhakar Prasad}
\email{prasa132@umn.edu}
\affiliation{School of Physics
and Astronomy, University of Minnesota, Minneapolis, MN 55455}
\altaffiliation{Also in Department of Physics and Astronomy, University of New Mexico, Albuquerque, NM 87131}

\date{\today}

\pacs{(100.6640) Superresolution; (110.3055) Information theoretical analysis;
 (110.7348) Wavefront encoding; (110.1758)
Computational imaging; (270.5585) Quantum information and processing}

\begin{abstract}

We calculate the quantum Fisher information (QFI) for estimating, using a circular imaging aperture, the length of a uniformly bright incoherent line source with a fixed mid-point and the radius of a uniformly bright incoherent disk shaped source with a fixed center. Prolate spheroidal wavefunctions (PSWFs) on a centered line segment and its generalized version on a centered disk furnish the respective bases for computing the eigenstates and eigenvalues of the one-photon density operator, from which we subsequently calculate QFI with respect to the spatial parameters of the two sources. Zernike polynomials provide a good set into which to project the full source wavefront, and such classical wavefront projection data can realize quantum limited estimation error bound in each case. We subsequently generalize our approach to analyze sources of arbitrary brightness distributions and shapes using a certain class of Bessel Fourier functions that are closely related to the PSWFs. We illustrate the general approach by computing QFI for estimating the lengths of the principal axes of a uniformly bright, centered elliptical disk.
\end{abstract}

\vspace{-1cm}


\maketitle

\section{Introduction}\label{sec:intro}
An extended incoherent source of a continuous irradiance distribution can be regarded as the limit in which the scale of spatial coherence on the source is comparable to the mean emission wavelength but small compared to the smallest spatial scale of change of its brightness distribution \cite{Goodman00}. Equivalently, it may be regarded as consisting of a collection of closely packed, equally bright point emitters emitting independently of one another such that their number density per unit area is proportional to the local source intensity $I(\br)$ and the total irradiance of an infinitesimal area element $dA$ centered at location $\br$  is $I(\br)\,dA$ in the continuous limit that the emitter spacing vanishes. It is possible to approximately evaluate the quantum Fisher information (QFI) matrix \cite{Toth14, Liu20}, whose inverse, the quantum Cram\'er-Rao bound (QCRB), yields the minimum variance with which one can estimate the size parameters of the source, by taking the continuous limit numerically. This was demonstrated recently \cite{Dutton19} for estimating the length of a uniformly bright, centered line source. 

The wholly numerical approach, however, misses the deeper insights afforded by a functional analysis that obviates such a discrete point-source representation altogether. We will use such a continuous functional analysis here to calculate the single-photon QFI with respect to (w.r.t) the size of the two simplest extended incoherent sources in one and two dimensions, namely the one-dimensional (1D) uniformly bright line source considered in Ref.~\cite{Dutton19} and the two-dimensional (2D) uniformly bright circular disk. As shown in Fig.~1, we will take both sources to be centered at a fixed origin, oriented transverse to the optical axis of a circular-aperture imager in the plane of Gaussian focus, and emitting monochromatically.  The 1D prolate spheroidal wave functions (PSWFs) \cite{Slepian61} and their generalized 2D versions \cite{Slepian64}, as we will show, furnish excellent bases in which to calculate the eigenstates and eigenvalues of the single-photon density operator (SPDO) and from them QFI for estimating the sizes for these two sources rather efficiently.
The ratio of the source length to the characteristic Airy diffraction width, we will see, determines the space-bandwidth parameter (SBP) of the corresponding PSWF problem, which serves as an effective dimensionality of the phase space of the photon density operator for the continuous line-source problem.
For the disk-source superresolution problem too, a similar SBP interpretation applies to the disk radius. 
\begin{figure}[htb]
\centering
\subfloat[]
{\includegraphics[width=0.45\columnwidth]{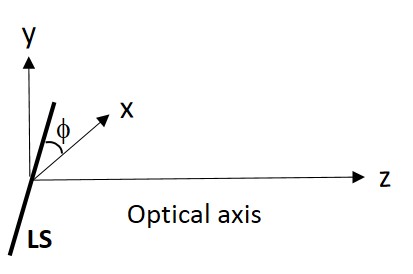}}
\subfloat[]
{\includegraphics[width=0.45\columnwidth]{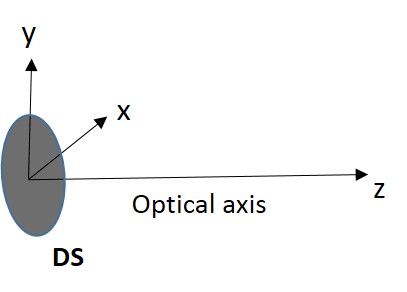}}\\
\caption{\label{fig:f1} (a) Centered, uniform line source (LS) in the plane of best focus transverse to the optical axis of an imager; (b) same arrangement as in (a) but for a centered, uniform disk source.}
\end{figure}

For an extended incoherent source with a well behaved intensity distribution that admits unique characterization in terms of its moments, calculating bounds on the quantum and classical Fisher information measures w.r.t.~those moments has previously allowed treatment of more general non-uniformly bright incoherent sources \cite{Tsang17, Chrostowski17, Tsang19, Zhou19,Bonsma-Fisher19}. The approach used in the present paper too can be generalized for nonuniformly bright sources, as we will also discuss, so as an alternative approach it may be even more generally applicable in situations where such moments might not exist to all orders. We can also generalize the QFI calculation for the nonsymmetric-point-source-pair problem \cite{Prasad20a} to many point sources of unequal intensities and take the appropriate continuous limit in a manner analogous to the analysis in Ref.~\cite{Dutton19} for general nonuniformly bright 1D and 2D sources. Such a calculation would once again fail, however, to provide any useful physical insights based on the source brightness geometry.

We begin the paper by introducing expressions for SPDO for the two sources in terms of pure point-emission states and their wavefunctions in the imaging pupil. In Sec.~III, we note that the problem of calculating QFI for estimating the length of a centered, monochromatic line source can be mapped isomorphically to that for estimating the separation of a symmetric point-source pair with fixed center under uniform broadband emission that we considered recently \cite{Prasad20b}. As such, we simply recall many of the expressions already derived in Ref.~\cite{Prasad20b} and point out the modifications needed to complete the calculation of QFI w.r.t.~the line-source length. We then calculate QFI w.r.t.~the line-source orientation by making use of certain rigorous identities that the eigenstates of SPDO obey, while also showing that a joint estimation of the length and orientation of the line source incurs no additional penalty at the quantum level when compared to their independent estimatiom. We next show rigorously that classical projections of the imaging wavefront in the Zernike polynomial basis can yield classical Fisher information (CFI) that approaches QFI in the photon-counting limit of detection for extreme subdiffractive scales. 

In Sec.~IV, we treat the 2D problem of disk-radius estimation by calculating the corresponding QFI. As a useful by-product of this calculation, we will derive expressions for the 2D PSWFs that, as superpositions of ordinary Bessel functions, are automatically analytic functions over the entire semi-infinite radial line. 
In fact, as we will see, the Bessel-function-based approach works well to formulate radial PSWFs for spaces of arbitrary dimensionality. For the disk problem too, we show that Zernike projections yield CFI that approaches QFI in the limit of small radius. Such projections thus might provide an optimal basis for achieving quantum limited superresolution imaging for both sources.

In the final long section of the paper, we generalize our approach to sources of arbitrary shapes and brightness distributions. We show how the radial Bessel representation of the 2D PSWFs, which we develop in an Appendix, is particularly useful for tackling the general brightness distribution, since unlike the one based on radial Zernike polynomials \cite{Slepian64}, the radial Bessel form remains valid everywhere in the infinite 2D plane. Further, it can be combined with the Fourier angular representation to extend it readily to radially non-symmetric situations as well. We illustrate this generalized approach by calculating QFI for estimating the lengths of the principal axes of a uniformly bright elliptical disk shaped source of {\em a priori} known center and orientation.

\section{The Single-Photon Density Operator for the Uniform Line and Disk Sources}
A single photon emitted by centered, uniformly bright, incoherent line and disk-shaped sources and then transmitted into an imaging system with aperture function $P(\bu)$ may be described, respectively, by the following SPDOs:
\ba
\label{rho12}
\hrho_1 = &{1\over l}\int_{\cL} dx\, |K_{x\hat\bl}\ra\la K_{x\hat \bl}|;\nn
\hrho_2 = &{1\over \pi a^2}\int_{\cD_a} dA\, |K_\br\ra\la K_\br|,
\end{align}
in which $\cL$ denotes the symmetric interval $(-l/2,l/2)$ along the line-source length vector, $\bl=l\hat\bl$, and $\cD_a$ the interior of the disk source of radius $a$. The symbol $|K_{x\hat\bl}\ra$ denotes the state vector of an imaging photon when emitted by the line-source element centered at position $x\hat\bl$ and $|K_\br\ra$ similarly that of an imaging photon emitted by the disk-source area element centered at point $\br$. The respective wavefunctions have the following form in the system's exit pupil \cite{Goodman96}:   
\ba
\label{wavefunction}
\la \bu| K_{x\hat\bl}\ra = & {1\over\sqrt{\pi}}P(\bu)\, \exp(-i2\pi x \hat\bl\cdot\bu);\nn 
\la \bu| K_\br\ra = & {1\over\sqrt{\pi}}P(\bu)\, \exp(-i2\pi \br\cdot\bu), 
\end{align}
where the normalized pupil position vector $\bu$ is the true pupil position vector divided by the radius $R$ of the circular, clear exit pupil. The pupil function, $P(\bu)$, takes the value 1 inside the unit disk centered at the origin and vanishes elsewhere. For the line source, $x$ denotes the position along the source in normalized image-plane coordinates obtained by dividing the true physical position $s$ by the characteristic Airy diffraction scale corresponding to the center optical wavelength, $\lambda_0$, and the distance, $z_I$, of the image plane from the exit pupil, {\em i.e.,}  $x=s/(\lambda_0 z_I/R)$. For the disk source, $\br$ denotes the transverse (2D) location vector of a point on the disk in normalized image-plane coordinates,  obtained similarly by dividing the physical position vector $\bs$ by the same diffraction scale,  $\br=\bs/(\lambda_0 z_I/R)$.  

The SPDOs (\ref{rho12}) may be expressed more simply in further scaled integration variables, $x\to lx$, $\br\to a\br$, as
\ba
\label{srho12}
\hrho_1 = &\int_{-1/2}^{1/2} dx\, |K_{x\bl}\ra\la K_{x\bl}|;\nn
\hrho_2 = &{1\over \pi}\int_{\cD_1} dA\, |K_{a\br}\ra\la K_{a\br}|,
\end{align}
where $\cD_1$ is the unit-radius disk centered at the origin. In the scaled variables, wavefunctions (\ref{wavefunction}) take the form,
\ba
\label{swavefunction}
\la \bu| K_{x\bl}\ra = & {1\over\sqrt{\pi}}P(\bu)\, \exp(-i2\pi x lu_l),\ \ x\in (-1/2,1/2);\nn 
\la \bu| K_{a\br}\ra = & {1\over\sqrt{\pi}}P(\bu)\, \exp(-i2\pi a\br\cdot\bu),\ \ \br\in\cD_1,
\end{align}
in which $u_l=\hat\bl\cdot\bu=u\cos(\phi-\phi_u)$ is the projection of vector $\bu$ on the line source.

The pure-state wavefunctions corresponding to two different source points have a non-vanishing overlap integral,
\ba
\label{overlap}
O_a(\br-\brp) &\defeq \la K_{a\br}|K_{a\brp}\ra \nn
&= {1\over \pi}\int d^2 u |P(\bu)|^2\exp [i2\pi a(\br-\brp)\cdot\bu],
\end{align}
with $a$, $\br$, and $\brp$ for the disk source replaced by $l$, $x\,\hat \bl$, and $x'\,\hat \bl$, respectively, for the line source. For a circular unit-radius clear pupil, for which $P(\bu)$ is simply its indicator function, the above integral evaluates rigorously to a form involving the Bessel function $J_1$,
\be
\label{overlap1}
O_a(\br-\br') = {J_1(2\pi a|\br-\brp|)\over \pi a|\br-\br'|},
\ee
which reduces to 1 when $\br=\brp$, as required by the normalization of the single-photon pure states and of the SPDOs (\ref{srho12}), namely $\Tr\hrho_1=\Tr\hrho_2=1$.

\section{Length and Orientation Estimation for a Uniformly Bright, Centered Line Source}

The evaluation of QFI for the uniformly bright, centered line source is greatly faciliated by its correspondence with the 2D localization problem of a point source that emits uniformly in a finite bandwidth, a problem we have recently discussed \cite{Prasad20b}. 

\subsection{Isomorphism between  Finite-Bandwidth Pair Separation and Monochromatic Line-Source Estimation Problems} 

The wavefunction for the pure state $|K_{Bf}\ra$ of a single photon emitted at frequency $\omega_0(1+Bf)$ by a point source located at position $ l\hat l$ takes the form \cite{Prasad20b}
\ba
\label{fwavefunction}
\la\bu|K_{Bf}\ra = &{1\over\sqrt{\pi}}P(\bu)\, \exp[-i2\pi (f+1/B)Blu_l],\nn
                             & \quad f\in (-1/2,1/2).
\end{align}
A comparison of  Eq.~(\ref{fwavefunction}) and the first of Eqs.~(\ref{swavefunction}) suggests a formal correspondence between the finite-bandwidth point-source and monochromatic line-source problems under the mapping, $ f+1/B\to x$, $Bl\to l$. In the language used to describe the 1D prolate spheroidal wavefunctions \cite{Slepian61}, the space-bandwidth product (SBP) parameter, $C=\pi Bl$, for the former problem must be replaced by $C=\pi l$ for the latter problem. With these correspondences, the two problems become formally identical. Applying these mappings, we can therefore calculate the set of SPDO eigenvalues $\{\lambda_i,\ i=1,2,\ldots\}$ and eigenstates $\{|\lambda_i\ra,\ i=1,2,\ldots\}$ and from them quantum Fisher information (QFI) w.r.t. $l$ by using essentially the same computer codes as those used for the finite-bandwidth source-localization problem \cite{Prasad20b}. A full description of the solution of the finite-bandwidth problem, including the detailed evaluation of QFI, can be found in that reference, which we do not present here to avoid duplication. For the purpose of calculating the QFI matrix for joint estimation of the length and orientation of the line source, we do, however, list a few important properties of the eigenvalues and eigenstates and also derive a few rigorous relations involving them.

\subsubsection{Eigenstate Expansion}
We may express the eigenstates of SPDO in terms of pure point-emission photon states as
\be
\label{expansion1}
|\lambda_i\ra = \int_{-1/2}^{1/2} dx\, C_i(x) |K_{x\bl }\ra,
\ee
in which $\lambda_i$ is a non-zero eigenvalue of $\hrho_1$ given in Eq.~(\ref{srho12}).
The coefficient function, $C_i(x)$, obeys the equation,
\be
\label{coeff_eq1}
\int_{-1/2}^{1/2} dx' C_i(x')\, O_l(x-x')= \lambda_i C_i(x), \ \ x\in (-1/2,1/2).
\ee
It can be chosen to be a real function due to the reality of the overlap function, 
\be
\label{overlap_line}
O_l(x-x')\defeq \la K_{x\bl}|K_{x'\bl}\ra= {J_1(2\pi l(x-x'))\over \pi l(x-x')},
\ee
which is obtained from expression (\ref{overlap1}) by applying the correspondences, $a\to l,\ \br\to x\bl,\ \brp\to x'\bl$.

\subsubsection{Eigenstate Overlap with a Point-Emission State}
The relation,
\be
\label{overlap_eigenstate1}
\la K_{x\bl}|\lambda_i\ra = \lambda_i C_i(x),
\ee
can be derived easily by using expansion (\ref{expansion1}), taking its inner product with state $|K_{x\bl}\ra$, and using the integral equation (\ref{coeff_eq1}) obeyed by the coefficient function $C_i(x)$. An analogous relation holds for the eigenstates belonging to the null space of $\hrho_1$ as well,
\be
\label{overlap_eigenstate_null1}
\la K_{x\bl}|\psi_j\ra=0,\ \ \{|\psi_j\ra\, \mid \, \hrho_1|\psi_j\ra=0\}.
\ee
 All eigenstates can be - and are - chosen to be orthonormal, $\la\lambda_i|\lambda_j\ra=\delta_{ij}$, $\la\lambda_i|\psi_j\ra=0$, $\la\psi_j|\psi_k\ra=\delta_{jk}.$

\subsubsection{Certain Identities involving Eigenvalues and Eigenstates}
Based on properties (\ref{expansion1})-(\ref{overlap_eigenstate_null1}), we may derive some rigorous relations involving the eigenvalues, $\lambda_i$, and coefficient functions, $C_i(x)$. Consider the functions,
\be
\label{In}
I_n(x,x') \defeq \sum_i \lambda_i^n C_i(x)\,C_i(x'),\ \ n=1,2,\ldots
\ee
Use of relations (\ref{overlap_eigenstate1}) and (\ref{overlap_eigenstate_null1}), along with the reality of $C_i(x)$, allows us to evaluate $I_2$,
\ba
\label{I2}
I_2(x,x') =&\la K_{x\bl}|\left[\sum_{i\in \cR}|\lambda_i\ra\la\lambda_i|+\sum_{j\in{\cal N}}|\psi_j\ra\la\psi_j|\right]|K_{x'\bl}\ra\nn
=&\la K_{x\bl}|K_{x'\bl}\ra\nn
=&O_l(x-x'),
\end{align}
in which we used the completeness of all orthonormal eigenstates of $\hrho_1$ to reach the second equality. The symbols $\cR$ and ${\cal N}$ denote the sets of values of an index that labels the eigenstates of $\hrho_1$ in its range and null spaces, respectively. 

A recursive relation between $I_n$ and $I_{n+1}$ for an arbitrary positive integer $n$ follows immediately from a use of relations (\ref{In}) and  (\ref{coeff_eq1}),
\be
\label{In2Inp1}
\int_{-1/2}^{1/2} dx' I_n(x,x')\, O_l(x^{\prime\prime}-x')=I_{n+1}(x-x^{\prime\prime}).
\ee
For the special case of $n=1$, use of definition (\ref{overlap_line}) of the overlap function and equality (\ref{I2}) allows us to rewrite Eq.~(\ref{In2Inp1}) as
\be
\label{I1I2}
\int_{-1/2}^{1/2} dx' I_1(x,x')\,\la K_{\xpp\bl}|K_{x'\bl}\ra=\la K_{x^{\prime\prime}\bl}|K_{x\bl}\ra,
\ee
from which follows the more general relation between point-emission state vectors, 
\be
\label{I1state}
\int_{-1/2}^{1/2} dx' I_1(x,x')\,|K_{x'\bl}\ra=|K_{x\bl}\ra,
\ee
since $|K_{\xpp\bl}\ra$ are linearly independent states. In view of definition (\ref{In}), this may also be expressed as a sum rule,
\be
\label{I1state1}
\sum_i \lambda_i C_i(x) \int_{-1/2}^{1/2} dx' C_i(x')\,|K_{x'\bl}\ra=|K_{x\bl}\ra,
\ee
valid for all $x\in (-1/2,1/2)$.


\subsection{QFI for Joint Estimation of Length and Orientation of the Line Source}

For a simultaneous estimation of both the source length and orientation, the lowest possible variances are gievn by the diagonal elements of the inverse of the $2\times 2$ QFI matrix,
\be
\label{JointQFI1}
\bH=\left(
\begin{array}{ll}
H_{ll}& H_{l\phi}\\
H_{l\phi} & H_{\phi\phi}
\end{array}
\right),
\ee
where $H_{ll}$ is QFI for length estimation alone, which we have already evaluated by invoking the isomorphism discussed in Sec.~III A, $H_{\phi\phi}$ is QFI for orientation estimation alone, and $H_{l\phi}$ represents the effect of mutual interference of the two estimations in which each parameter serves as a nuisance parameter \cite{Berger99} for the other. The QFI matrix elements per photon are defined as the real part, denoted by symbol Re, of a trace, denoted by symbol Tr,
\be
\label{Hmunu}
H_{\mu\nu} = {\rm Re}\left[\Tr (\hrho_1 \hat L_\mu\hat L_\nu)\right],\ \ (\mu,\nu)=(l,\phi) \ {\rm or}\ (\phi,\phi),
\ee
in which $\hat L_\mu$ denotes the symmetric logarithmic derivative of $\hrho_1$ w.r.t.~parameter labeled by index $\mu$. As we showed in Ref.~\cite{YuPrasad18}, we may express $H_{\mu\nu}$ in terms of matrix elements of ordinary partial derivatives of $\hrho_1$ as
\begin{align}
\label{Hmunu1}
&H_{\mu\nu}={\rm Re}\left(\sum_{i\in \cR}{4\over \lambda_i}\langle \lambda_i|\partial_\mu \hat\rho_1\partial_\nu \hat\rho_1|\lambda_i\rangle\right)\nn
&+\sum_{i,j\in \cR}\left[{4\lambda_i\over {(\lambda_i+\lambda_j)}^2}-{4\over \lambda_i}\right]{\rm Re}[\langle \lambda_i|\partial_\mu \hat\rho_1|\lambda_j\rangle\langle \lambda_j|\partial_\nu \hat\rho_1|\lambda_i\rangle].
\end{align}

Using the product rule for derivatives on expression (\ref{srho12}), we may express $\pmu\hrho_1$ as
\be
\label{dmu_rho1}
\partial_\mu\hrho_1= \int_{-1/2}^{1/2}dx\, \left(\pmu|K_{x\bl}\ra\la K_{x\bl}|+|K_{x\bl}\ra\pmu\la K_{x\bl}|\right)
\ee
and thus a general matrix element of $\pmu\hrho_1$ as the integral,
\ba
\label{MatEl_rho1}
\la \lambda_i|\partial_\mu\hrho_1|\lambda_j\ra=&\int_{-1/2}^{1/2}dx\, \big[\la\lambda_i|\pmu|K_{x\bl}\ra\la K_{x\bl}|\lambda_j\ra\nn
&\qquad\qquad+\la\lambda_i|K_{x\bl}\ra\la\lambda_j\pmu| K_{x\bl}\ra^*\big]\nn
=&\int_{-1/2}^{1/2}dx\, \big[\lambda_j C_j(x)\la\lambda_i|\pmu|K_{x\bl}\ra\nn
&\qquad\qquad+\lambda_i C_i(x)\la\lambda_j\pmu| K_{x\bl}\ra^*\big],
\end{align}
in which we used relation (\ref{overlap_eigenstate1}) to reach the second equality. For the specific case of $\mu$ labeling the orientation parameter $\phi$, which is the angle that the line source makes with the $x$ axis in the transverse plane (see Fig.~1), we may use expansion (\ref{expansion1}) of an eigenstate to express the first matrix element on the right-hand side (RHS) of Eq.~(\ref{MatEl_rho1}) as 
\be
\label{MatEl_lambda_dphi_K}
\la\lambda_i|\partial_\phi|K_{x\bl}\ra=\int_{-1/2}^{1/2} dx' C_i(x') \la K_{x'\bl}|\partial_\phi|K_{x\bl}\ra.
\ee
Use of the first of the wavefunctions (\ref{swavefunction}) allows us to write the matrix element inside the $x'$ integrand in Eq.~(\ref{MatEl_lambda_dphi_K}) as the following-pupil plane integral:
\ba
\label{KdphiK}
\la K_{x'\bl}|\partial_\phi|K_{x\bl}\ra=&{i2\pi xl\over\pi}\int d^2 u\, P(\bu)\, u\sin(\phi-\phi_u)\nn
                                                                                     &\times \exp[i2\pi(x'-x) l u\cos(\phi-\phi_u)],
\end{align}
which vanishes for any radially symmetric pupil, like the clear circular pupil, due to its reflection symmetry about any radial direction, specifically under $\phi_u-\phi \to -(\phi_u-\phi)$ for which the integrand only changes sign without any other change. As a consequence, both expressions (\ref{MatEl_lambda_dphi_K}) and (\ref{MatEl_rho1}) must vanish identically for a radially symmetric pupil,
\be
\label{MatEl1}
\la\lambda_i|\partial_\phi|K_{x\bl}\ra=0,\ \la \lambda_i|\partial_\phi\hrho_1|\lambda_j\ra=0.
\ee
Use of the second of results (\ref{MatEl1}) in expression (\ref{Hmunu1}) greatly simplifies the latter whenever one or both of the indices $\mu,\nu$ refer to the orientation angle $\phi$,
\be
\label{Hlphi}
H_{\mu\nu} = \sum_{i\in \cR}{4\over \lambda_i}{\rm Re}\langle \lambda_i|\partial_\mu \hat\rho_1\partial_\nu \hat\rho_1|\lambda_i\rangle,\ (\mu,\nu)=(l,\phi)\ {\rm or} \ (\phi,\phi).
\ee

For $\nu=\phi$, we may evaluate Eq.~(\ref{Hlphi}) by taking a product of expression (\ref{dmu_rho1}) evaluated for $\mu=l$ and $\partial_\phi\hrho_1$, and then taking the diagonal matrix element of the product in state $|\lambda_i\ra$. A subsequent use of the second of the identities of Eq.~(\ref{MatEl1}) allows one to express $H_{l\phi}$ as
\be
\label{Hlphi1}
H_{l\phi}=\sum_{i\in \cR}{4\over \lambda_i}{\rm Re}\int_{-1/2}^{1/2} dx \la\lambda_i|K_{x\bl}\ra\partial_l\la K_{x\bl}|\partial_\phi\hrho_1|\lambda_i\ra.
\ee
Since the wavefunction corresponding to the state $\partial_l|K_{x\bl}\ra$ has the same parity as that of state $|K_{x\bl}\ra$ under reflection in the line source, $\phi_u-\phi\to -(\phi_u-\phi)$, we see by means of arguments similar to those used to prove the second equality in Eq.~(\ref{MatEl1}) why the matrix element inside the integrand of Eq.~(\ref{Hlphi1}) must vanish too,
\be
\label{dlKdphi_rho}
 \partial_l\la K_{x\bl}|\partial_\phi\hrho_1|\lambda_i\ra=0,
 \ee
and so
\be
\label{Hlphi2}
H_{l\phi}=0.
\ee
Rather fundamentally, thus, $l$ and $\phi$ can be estimated jointly without incurring any penalty of mutual interference of the two parameters. 

Analogous to the manner in which we derived Eq.~(\ref{Hlphi1}) from the more general Eq.~(\ref{Hlphi}), we may express QFI for estimating the orientation alone, $H_{\phi\phi}$, as
\be
\label{Hphi2}
H_{\phi\phi}=\sum_i {4\over \lambda_i}\int_{-1/2}^{1/2} dx \la\lambda_i|K_{x\bl}\ra\partial_\phi\la K_{x\bl}|\partial_\phi\hrho_1|\lambda_i\ra,
\ee
where and henceforth we omit any reference to the index set $\cR$ as being understood without having to be explicitly stated. If we now set $\mu=\phi$ in expression (\ref{dmu_rho1}) and note that $\partial_\phi\la K_{x\bl}|\lambda_i\ra$, being merely the complex conjugate of the first identity in Eq.~(\ref{MatEl1}), vanishes, we may simplify expression (\ref{Hphi2}),
\ba
\label{Hphi2a}
H_{\phi\phi}&=\sum_i{4\over \lambda_i}\iint_{-1/2}^{1/2} dx \, dx'\la\lambda_i|K_{x\bl}\ra\partial_\phi\la K_{x\bl}|\partial_\phi|K_{x'\bl}\ra\nn
&\qquad\qquad\qquad\qquad\times\la K_{x'\bl}|\lambda_i\ra\nn
                         =&4\sum_i \lambda_i\iint_{-1/2}^{1/2} dx \, dx'C_i(x)\, C_i(x')\partial_\phi\la K_{x\bl}|\partial_\phi|K_{x'\bl}\ra\nn
                         =&4\iint_{-1/2}^{1/2} dx \, dx'I_1(x,x')\partial_\phi\la K_{x\bl}|\partial_\phi|K_{x'\bl}\ra
\end{align}
in which the second equality follows from the first upon using identity (\ref{overlap_eigenstate1}) twice and the reality of the coefficient function $C_i(x)$, and the final equality simply replaces the sum over $i$ by $I_1(x,x')$ defined via Eq.~(\ref{In}).

To further evaluate expression (\ref{Hphi2a}), we first differentiate relation (\ref{I1state}) w.r.t.~$\phi$. Since neither the eigenvalues $\lambda_i$ nor the coefficient functions $C_i(x)$ depend on the orientation of the line source, as Eq.~(\ref{coeff_eq1}) that determines them shows, this differentiation yields the relation,
\be
\label{dI1state}
\int_{-1/2}^{1/2} dx' I_1(x,x')\,\partial_\phi|K_{x'\bl}\ra=\partial_\phi|K_{x\bl}\ra.
\ee
Computing the inner product of this relation with the state $\partial_\phi|K_{x\bl}\ra$ and integrating the result over $x$ in the interval $(-1/2,1/2)$ evaluates expression (\ref{Hphi2a}) for orientational QFI as
\be
\label{Hphi2b}
H_{\phi\phi}=4\int_{-1/2}^{1/2} dx\,\partial_\phi\la K_{x\bl}|\partial_\phi|K_{x\bl}\ra.
\ee
Using the first of the forms (\ref{swavefunction})  for the wavefunction, we can now fully evaluate expression (\ref{Hphi2b}) for $H_{\phi\phi}$ for a clear circular pupil of unit scaled radius as 
\ba
\label{Hphi2c}
H_{\phi\phi}=&{4\over \pi}\int_{-1/2}^{1/2} dx\,(2\pi xl)^2\int d^2 u P(u)\,u^2 \sin^2(\phi_u-\phi)\nn
                         =&{\pi^2 \over 3}l^2.
\end{align}

Equation (\ref{Hphi2c}) is an important result of the present paper. It implies that the minimum variance for an unbiased estimation of its angular orientation, $\phi$, is inversely proportional to that squared length. For a centered line source, if its length $l$ were fixed and perfectly known {\it a priori}, the estimation of the spatial position of its extreme points on the circle of radius $l/2$ that they must lie on would correspondingly have a minimum variance equal to $(l/2)^2$ times that for estimating $\phi$, namely $(l/2)^2/H_{\phi\phi}$, which evaluates to a constant $3/(2\pi)^2$. The inverse square law behavior of the minimum variance of angular-orientation estimation w.r.t.~$l$ reflects the increasing difficulty of determining the orientation of a uniformly lit line source with decreasing length, with such estimation becoming fundamentally intractable for $l\lesssim \sqrt{3}/\pi=0.55$.  

\subsection{Zernike Projections and Joint Estimation of Source Length and Orientation}
Before presenting results of a numerical evaluation of QFI, $H_{ll}$, for estimating the source length, we discuss how mutually orthogonal Zernike-mode projections that we previously demonstrated \cite{YuPrasad18,PrasadYu19,Prasad20b} as attaining QFI for the point-source localization and separation problems can do the same for the mononchromatic, centered line-source problem as well. As noted earlier \cite{Prasad20a}, Zernikes might constitute an optimal set of projection modes for all superresolution imaging of uniformly bright sources when using a clear-circular-pupil imager.

We first consider projection data for a small number of low-order Zernikes. The first four of them, in Noll's single-index scheme \cite{Noll76}, are defined as the following functions of polar coordinates over the unit disk, $0\leq u <1$, in the pupil plane:
\ba
\label{Z1234}
Z_1(\bu)=&{1\over \sqrt{\pi}};\ \ 
Z_2(\bu)={2\over\sqrt{\pi}}u\,\cos\phi_u;\nn 
Z_3(\bu)=&{2\over\sqrt{\pi}}u\,\sin\phi_u;\ \ 
Z_4(\bu)=\sqrt{3\over \pi}(2u^2-1),
\end{align}
with each having unit norm over the unit disk, $\la Z_i|Z_i\ra=1$. In applications involving their use to represent wavefront phase over a circular pupil, they are known as the piston, tip, tilt, and defocus modes, respectively. In view of expression (\ref{srho12}) for the line-source SPDO $\hrho_1$, we may express the probability of observing a single photon in the $j$th Zernike mode, $P_j=\la Z_j|\hrho_1|Z_j\ra$, as
\be
\label{ProbZj}
P_j={1\over \pi}\int_{-1/2}^{1/2} dx \left\vert\int P(\bu)\exp(-i2\pi x\, l\,u_l)\, Z_j(\bu) d^2u\right\vert^2,
\ee 
in which we used the first of expressions (\ref{swavefunction}) for the wavefunction, $\la \bu|K_{x\bl}\ra$. These probability integrals are evaluated in Appendix A for arbitrary values of mode index $j$. 

The classical Fisher information (CFI) \cite{VT68} for estimating $(l,\phi)$ from $N$ projections is the $2\times 2$ symmetric real matrix,
\be
\label{JointCFI1}
\bF^{(N)}=\left(
\begin{array}{ll}
F_{ll}^{(N)}& F_{l\phi}^{(N)}\\
F_{l\phi}^{(N)} & F_{\phi\phi}^{(N)}
\end{array}
\right),
\ee
where $F_{ll}^{(N)}$ is CFI for length estimation alone, $F_{\phi\phi}^{(N)}$ is CFI for orientation estimation alone, and $F_{l\phi}^{(N)}$ represents the effect of interference of the two parameters whereby each parameter serves as a nuisance parameter for the other in their joint estimation. The three nontrivial matrix elements in Eq.~(\ref{JointCFI1}) are defined per photon in terms of the set of single-photon probabilities, $\{P_1,\ldots,P_N\}$, by the relation,
\be
\label{Fmunu}
 F_{\mu\nu}^{(N)}=\sum_{j=1}^N {\pmu P_j\ \pnu P_j\over P_j}+ {\pmu \bar P\ \pnu \bar P\over \bar P},\ \ \mu,\nu=l,\phi,
\ee
in which the last term, with $\bar P=1-\sum_{j=1}^N P_j$, represents the contribution of the \emph{unobserved} modes. 

By including further Zernike modes beyond the first four into our projection data, we may improve CFI continually and push it closer to QFI. But can one truly reach QFI by including {\em all} Zernike modes? We answer this question by setting $N$ equal to $\infty$ in expression (\ref{Fmunu}).
As we show in detail in Appendix A, we may express the three different matrix elements of CFI, $\bF^{(\infty)}$, when all Zernike mode projections are included as
 \ba
\label{Ffull}
F_{ll}^{(\infty)} = & {1\over l^2}\Bigg[{4\over (\pi l)^3}\sum_{p=0}^\infty(p+1)^2{J_{p+1}^4(\pi l)\over \int_0^{\pi l} dw J_{p+1}^2(w)/w^2}-1\Bigg];\nn
F_{\phi\phi}^{(\infty)}=&{16\over \pi l}\Bigg[\int_0^{\pi l}dw \left({1\over 4}-{J_{1}^2(w)\over w^2}\right)\nn
                    &-\sum_{p=2,4,\ldots} (p+1)\int_0^{\pi l}dw {J_{p+1}^2(w)\over w^2}\Bigg];\nn
F_{l\phi}^{(\infty)}=&0.                    
\end{align}
The vanishing of the off-diagonal matrix elements of CFI, like QFI, indicates the absence of any mutual intereference of the two parameters, $l$ and $\phi$, in their joint estimation using Zernike mode projections. The two can be estimated independently to minimum variances that are given by the reciprocals of the diagonal elements of CFI.

\begin{figure}
\centerline{
\includegraphics[width=0.5\textwidth]{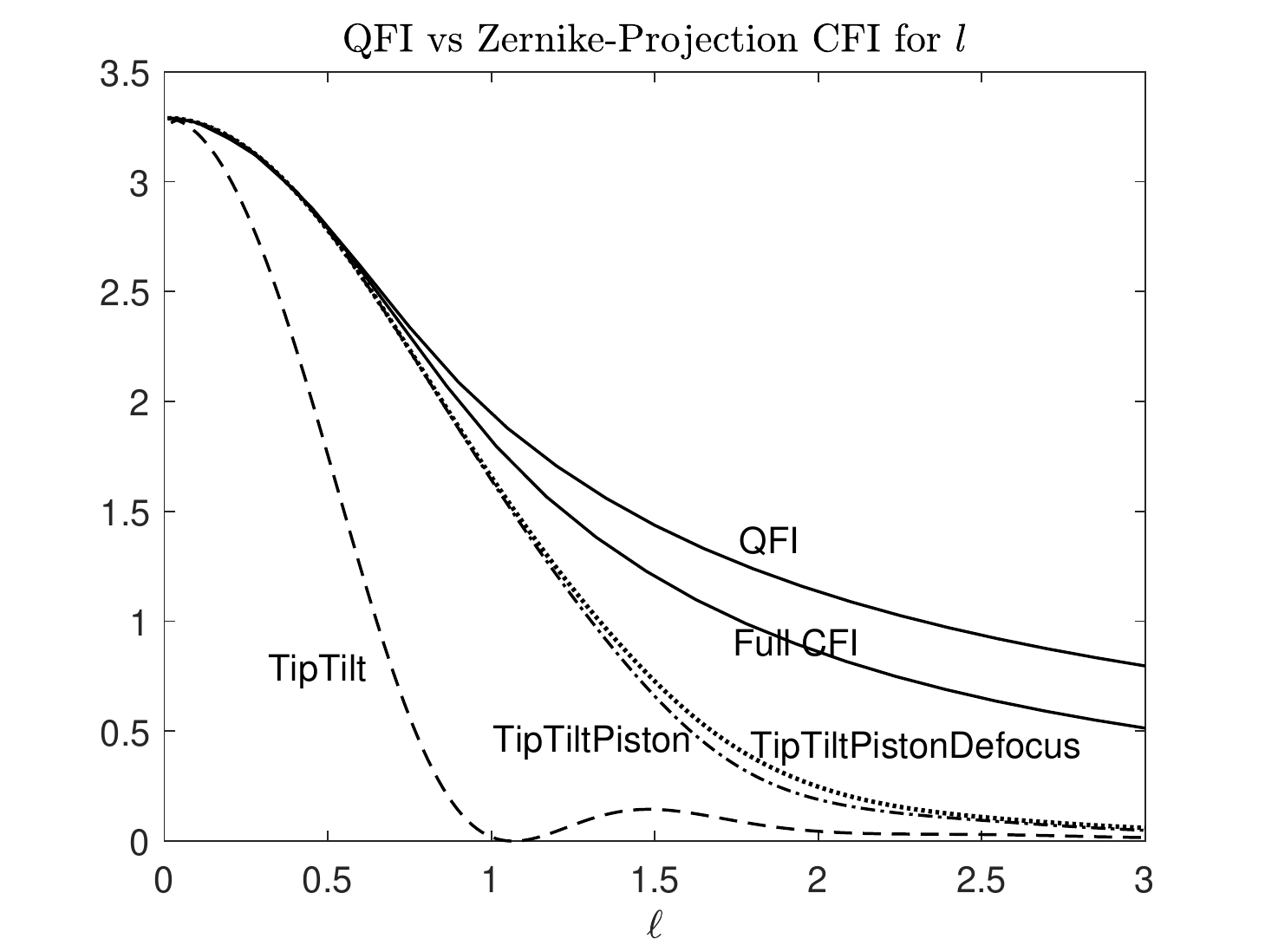}}
\caption{QFI and CFI for estimating the line-source length, $l$. The upper and lower solid curves display numerically computed values of QFI and full Zernike-based CFI, while the dashed, dash-dotted, and dotted curves display finite-mode CFI resulting from only observing either Zernike tip-tilt or tip-tilt-piston or tip-tilt-piston-defocus projections, respectively.}
\end{figure}
In Fig.~2, we use a solid curve to join the values of QFI for estimating the length, $l$, of the monochromatic, uniformly-bright, centered line source, obtained for a number of discrete values of $l$ by the method outlined in Sec.~III.A. The values of $F_{ll}^{(2)}$ when projections into only the tip-tilt modes $Z_2$ and $Z_3$ are observed are shown by the dashed curve. Note the convergence of these two curves as $l\to 0$, indicating that the tip-tilt modes are matched filters for the length coordinate of the source. When the contribution of the piston Zernike, $Z_1$, is added to CFI, the overall CFI, as shown by the dash-dotted curve, improves quite dramatically, particularly for values of $l\leq 1$. Adding the contribution of the defocus Zernike, $Z_4$, seems to make a discernible difference only in the range, $1.25< l < 2.25$.

Numerically computed values of expression (\ref{Ffull}) for the full CFI, $F_{ll}^{(\infty)}$, when all Zernike mode projections are included, is shown by the second curve from the top in Fig.~2. It still falls short of the ultimate upper bound, the corresponding QFI, $H_{ll}$, plotted as the uppermost curve in this figure, with the absolute gap between QFI and full CFI increasing with increasing source length.  The persistence of the finite QFI-CFI gap, although greatly reduced when compared to the case when only the lowest four Zernikes are included, may indicate the unattainability of QFI by {\em any} measurement that can be made on the line source to estimate its length.

We next plot results for QFI and CFI for estimating the line-source orientation. For a sub-diffractive source length, $l<<1/\pi$, by approximating $1/4-J_1^2(w)/w^2$ by its small-argument value, $w^2/16$, while neglecting all other terms in the second expression of Eq.~(\ref{Ffull}), we may calculate the limiting value $\pi^2 l^2/3$ for $F_{\phi\phi}$, which is the same as QFI, $H_{\phi\phi}$, given by Eq.~(\ref{Hphi2c}). Note that the same limiting value is obtained when only the lowest pair of $m\neq 0$ Zernikes, namely the tip-tilt Zernilkes, $Z_2$ and $Z_3$, are included, corresponding to only the first term of the first sum in Eq.~(\ref{Fphiphi2}).  In other words, the two lowest-order Zernike projections, $Z_2$ and $Z_3$, that have finite sensitivity to source-orientation angle $\phi$ can already achieve QFI for estimating that angle in this limit, as we see from the lowest curve in Fig.~3. The gap between the QFI and CFI is increasingly reduced by including more and more angle-dependent Zernikes beyond $Z_2$ and $Z_3$.  However, as Fig.~3 shows, even when all Zernikes are included in the projection data, orientation-estimation CFI still falls rather short of its ultimate upper bound, $H_{\phi\phi}$. 

All four information measures, $H_{ll}$, $H_{\phi\phi}/l^2$, $F_{ll}$, and $F_{\phi\phi}/l^2$, which we have displayed in Figs.~2 and 3, converge to the same value, $\pi^2/3$, in the limit $l\to 0$. This important result affirms the equality of the minimum root-mean-squared errors with which both the longitudinal and transverse coordinates of the extremeties of the linear source can be estimated, independently, in this limit.     
\begin{figure}
\centerline{
\includegraphics[width=0.5\textwidth]{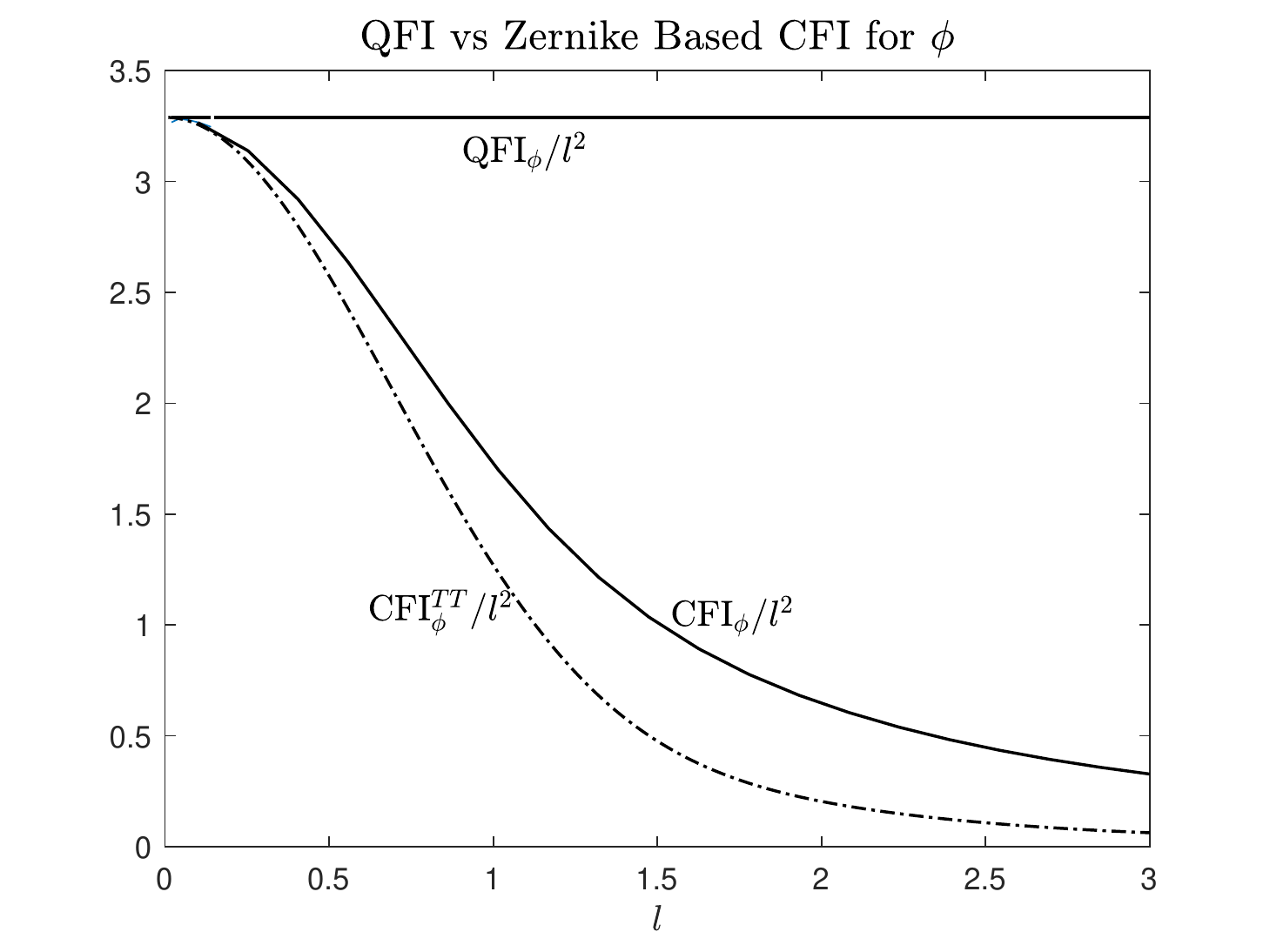}}
\caption{QFI and CFI for estimating the line-source orientation, $\phi$. The horizontal solid line displays the constant value, $\pi^2/3$, of $H_{\phi\phi}/l^2$, while the lowest of the curves is a plot of the corresponding CFI value, $F_{\phi\phi}/l^2$,  when only projections into the lowest two $\phi$ dependent Zernikes, namely tip and tilt, $Z_2$ and $Z_3$, are observed. The middle curve shows the increase of CFI resulting on including all Zernikes, as computed numerically from Eq.~(\ref{Fphiphi2}). }
\end{figure}

\subsection{Maximum-Likelihood Estimation of $l$ and $\phi$}
An effective practical approach for estimating the length and orientation of the source is based on the use of a maximum-likelihood (ML) algorithm. The probability of observing a set of counts $\{m_1,\ldots,m_N\}$ in $N$ projection modes, for a given pair of values of $l$ and $\phi$ and a given, fixed quantum efficiency, $\eta$, of the sensor pixels is given \cite{YuPrasad18} by a multinomial probability distribution (MPD) consisting of the product of a factor that depends on $\eta$ and a second factor that depends on the mode projection probabilities $\{P_1,\ldots,P_N\}$, with both factors also dependent on the observed counts. Since the ML estimator maximizes the probability of the observed counts w.r.t.~the parameters, $l$ and $\phi$, being estimated, we may for this purpose treat the first factor as an overall constant multiplier, $C$, and only keep the second factor that depends on those parameters through the dependence of the mode probabilities on them. The joint-count probability may thus be expressed as
\be
\label{CountProb}
{\rm Prob}(\{m_k\}|M) =C{{\bar P}^{\bar m}\over {\bar m}!}\left[\prod_{k=1}^N { P_k^{m_k}\over m_k!}\right],
\ee 
where $\bar m= M-\sum_{i=1}^N m_i$ is the number of counts {\em not observed} by the $N$ mode projections. Setting the first partials of the logarithm of Eq.~(\ref{CountProb})  w.r.t.~$l$ and $\phi$ to 0 generates ML estimates of the latter, independent of $\eta$, as solution of the following pair of equations:
\ba
\label{MLestimates}
\sum_{k=1}^N\left({m_i\over P_i}+{\bar m\over \bar P}\right){\partial P_i\over\partial l}=&0;\nn
\sum_{k=1}^N\left({m_i\over P_i}+{\bar m\over \bar P}\right){\partial P_i\over\partial \phi}=&0.
\end{align}

When wavefront projections into only the tip and tilt Zernike modes, $Z_2$ and $Z_3$, are used, since, as Eq.~(\ref{ProbZ1234}) shows, $P_2=P_3\cot^2\phi$, it follows from the first of Eqs.~(\ref{MLestimates}) that
\be
\label{MLestimates1}
\left[\left({m_2\over P_2}+{\bar m\over \bar P}\right)\cot^2\phi+\left( {m_3\over P_3}+{\bar m\over \bar P}\right)\right]{\partial P_3\over \partial l}=0.
\ee
Since the terms inside the square brackets in Eq.~(\ref{MLestimates1}) being positive cannot have a vanishing sum,  the condition,
\be
\label{l_estimate}
{\partial P_3\over \partial l}=0,
\ee
must hold. In view of expression (\ref{ProbZ1234}) for $P_3$, this is equivalent to the requirement,
\be
\label{l_estimate1}
{1\over\pi l}\int_0^{\pi l}dw{J_2^2(w)\over w^2 } = {J_2^2(\pi l)\over (\pi l)^2},
\ee
which is not only independent of the observed data but also cannot, in general, be satisfied even in the limit $l\to 0$. In other words, tip-tilt projections alone fail to yield an ML estimate for the source length.  
  
But that is not so for the estimate of the source orientation. From the second of the equations (\ref{MLestimates}), since $\partial P_2/\partial \phi=-\partial P_3/\partial\phi=-P_2\sin 2\phi$, does not vanish, in general, it follows that
\be
\label{MLphi}
{m_2\over P_2}+{\bar m\over \bar P}= {m_3\over P_3}+{\bar m\over \bar P},
\ee
yielding the ML estimate of the source orientation as
\be
\label{MLphi1}
\tan^2\phi = {P_3\over P_2}={m_3\over m_2}.
\ee
The ML estimates for both the source length and orientation get more constrained as more mode projections are observed.
 
\section{Radius Estimation for a Uniformly Bright, Fixed-Center Disk}
We now turn to the calculation of QFI and CFI for estimating the radius of a uniformly bright disk source with a fixed center chosen at the origin of the coordinate system. Computing QFI from Eq.~(\ref{Hmunu1}), in which $\hrho_1$ is replaced by $\hrho_2$ as being the appropriate SPDO for the uniformly bright disk, requires computing accurately the non-zero eigenvalues and eigenstates of SPDO first. 

\subsection{The SPDO Eigenvalue Problem}
Let $|\lambda\ra$ be an eigenstate of $\hrho_2$ with a non-zero eigenvalue $\lambda$. Since $\hrho_2$, given in scaled form by Eq.~(\ref{srho12}), is supported over the subspace $,\cH_a\defeq \{|K_{a\br}\ra \mid \br\in \cD_1\}$, all its eigenstates with non-zero eigenvalues may also be chosen to be fully contained in $\cH_a$. Consider the expansion of $|\lambda\ra$ in this basis,
\be
\label{expansion}
|\lambda\ra = {1\over \pi } \int_{\cD_1} dA \, \Cl(\br) |K_{a\br}\ra.
\ee
On substituting Eq.~(\ref{expansion}) and expression (\ref{srho12}) for $\hrho_2$ into the eigenstate relation,
\be
\label{eigenrelation2}
\hrho_2|\lambda \ra =\lambda |\lambda\ra,
\ee
and then equating the coefficients of each $|K_{a\br}\ra$ term on the two sides of the resulting equation, which is permitted due to the linear independence of these single-photon states that are emitted by different points on the incoherent source, we obtain the following integral equation for the coefficient function, $\Cl(\br)$:
\be
\label{eigenrelation2a}
{1\over \pi} \int_{\cD_1} O_a(|\br-\brp|)\, \Cl (\brp)\, dA' = \lambda \Cl(\br), \ \ \br\in\cD_1,
\ee
where the kernel function is given by Eq.~(\ref{overlap1}).  In view of Eq.~(\ref{eigenrelation2a}), we immediately recognize the coefficient functions $\Cl(\br)$, of which there are countably infinitely many, as being the 2D version of the orthogonal set of generalized PSWFs first discussed by Slepian \cite{Slepian64}, with the quantity, $c=2\pi a$, serving as the space-bandwidth parameter (SBP). Many of their properties were discovered by Slepian himself in that seminal paper, but a few subsequent papers \cite{GreengardSerkh18,Lederman16,Shkolnisky07}  have derived further exact and asymptotic properties and provided useful analytical and numerical bases for them, while reviewing and simplifying Slepian's analysis. Here we will simply list some of those properties that are immediately relevant to our problem. 

Since $O_a(|\br-\brp|)$ is invariant both under rotations about the origin and reflections in the $x$ axis, its complete orthonormal set of eigenfunctions in 2D polar coordinates, $\br=(r,\theta)$, may be expressed in the separable form,
\ba
\label{PSWF}
C_{Nn}^{(\pm)}(\br)=& \Psi_{Nn}(r) \mu_N\left\{
\begin{array} {l}
\cos N\theta\\
\sin N\theta,
\end{array}
\right.  \nn
&N=0,1,\ldots,\ \ n=1,2,\ldots,
\end{align}
in which the coefficient $\mu_N$, takes the values, $\mu_0=1/\sqrt{2\pi}$ and $\mu_N=1/\sqrt{\pi},$ $N\geq 1$, ensuring that the angular harmonics are properly normalized over the fundamental angular period of $2\pi$. The two angular dependences are denoted by the superscripts $\pm$, which indicate their parity under reflection in the $x$ axis, {\it i.e.,} under $\theta\to-\theta$, but neither the eigenvalues nor the radial functions $\Psi_{Nn}(r)$, as we shall presently see, depend on this parity. The latter may be shown \cite{Slepian64} to be eigenfunctions of the Bessel integral operator of order $N$ over the unit disk,
\be
\label{RadialIntEq}
\int_0^1 dr' r' J_N(crr')\Psi_{Nn}(r')  =\beta_{Nn}\Psi_{Nn}(r), \ 0\leq r\leq 1,
\ee
with the eigenvalues of interest, $\lambda_{Nn}$, being proportional to $\beta_{Nn}^2$.
These functions may be expressed as sums over normalized radial Zernike polynomials, $\bar R_{Nk}(r)$,
\be
\label{RadialPSWF}
\Psi_{Nn}(r)=\sum_{k=0}^\infty h_{Nnk} \bar R_{Nk}(r),\ 0\leq r\leq 1,
\ee
where the normalized Zernike polynomials are defined over the unit disk, $0\leq r \leq 1$, as 
\ba
\label{RadialZernike}
\bar R_{Nk}(r) = &\sqrt{2(N+2k+1)}\,r^N\sum_{m=0}^k (-1)^m{N+k\choose m}\nn
&\times{k\choose m} r^{2(k-m)}(1-r^2)^m.
\end{align}
The radial Zernikes (\ref{RadialZernike}) form an orthoronormal set along the radial dimension of the unit disk,
\be
\label{RadialNorm}
\int_0^1 dr\, r \bar R_{Nk}(r)\bar R_{Nk'}(r) =\delta_{kk'}.
\ee

\subsubsection{Some Properties of Eigenstates and Eigenvalues}

In light of the expansion (\ref{expansion}) for an eigenstate of SPDO in its range space, requiring such eigenstates to be orthonormal, {\it i.e.},  
\be
\label{EigenstateNorm}
\la\lambda_{Nn}^{(p)}|\lambda_{N'n'}^{(p')}\ra=\delta_{NN'} \delta_{nn'}\delta_{pp'},
\ee
amounts to requiring 
\ba
\label{EigenstateNorm1}
{1\over \pi^2} \iint_{D_1} dA\, dA'\, &C_{Nn}^{(p)}(\br)\, O_a(|\br-\brp|)\,C_{N'n'}^{(p')}(\brp)\nn
&=\delta_{NN'}\delta_{nn'}\delta_{pp'},
\end{align}
and thus, in view of eigenrelation (\ref{eigenrelation2a}), to requiring
\be
\label{EigenstateNorm2}
\int_{D_1} dA\, C_{Nn}^{(p)}(\br)\,C_{N'n'}^{(p')}(\br)={\pi\over \lambda_{Nn}}\delta_{NN'}\delta_{nn'}\delta_{pp'}.
\ee
Use of expansion (\ref{PSWF}) in this relation, followed by a use of the orthonormality of the angular parts of the eigenfunctions, yields the following radial-function orthogonality condition, valid for any $N$:
\be
\label{RadialEigenNorm}
\int_0^1 dr\, r\, \Psi_{Nn}(r)\,\Psi_{Nn'}(r)={\pi\over \lambda_{Nn}}\delta_{nn'}.
\ee
When expansion (\ref{RadialPSWF}) of the radial eigenfunction is substituted for each eigenfunction in Eq.~(\ref{RadialEigenNorm}) and orthormality (\ref{RadialNorm}) of the radial Zernike is used, we obtain the following eigenfunction orthogonality condition on the coefficients $h_{Nnk}$:
\be
\label{CoeffVectorNorm}
\sum_k h_{Nnk} h_{Nn'k}={\pi\over \lambda_{Nn}}\delta_{nn'}.
\ee

The eigenvalues $\lambda_{Nn}$, obtained by solving the integral equation (\ref{RadialIntEq}), evidently do not depend on the $x$-reflection parity, $p$, of the eigenfunctions $C_{Nn}^{(p)}(\br)$, and are thus doubly degenerate for $N>0$. For each value of the angular quantum number $N$, they are arranged in the decreasing order, with of order $(c-N-1)/2$ (for large $c$) of the eigenvalues being significantly different from zero for each value of $N$ of order 1. This mirrors the behavior of eigenvalues corresponding to the 1D PSWFs \cite{Fuchs64,Osipov12}. They also decrease with increasing values of $N$ even when the value of $n$ is of order 1. These trends are discussed more quantitatively in Appendix B.

The eigenvalues over all possible eigenfunctions of SPDO (\ref{rho12}) must add up to 1, as required by the normalization of the latter. In view of the double degeneracy of the eigenvalues corresponding to the eigenfunctions (\ref{PSWF}) w.r.t.~the sine and cosine angular dependences for $N >0$, and non-degenerate eigenvalues for $N=0$ (for which the $\sin N\theta$ dependent eigenfunction is not allowed), it follows that
\be
\label{eig_sum} 
\sum_{n=1}^\infty \lambda_{0n}+2\sum_{N=1}^\infty\sum_{n=1}^\infty \lambda_{Nn} = 1.
\ee
This sum rule serves as a check on our numerical calculations.

\subsubsection{Numerical Evaluation of the Eigenvalues and Eigenvectors}
Lederman \cite{Lederman16} has numerically computed the superposition coefficients $h_{Nnk}$ by diagonalizing a tridiagonal matrix (TDM), obtained by transforming \cite{Slepian64} the original integral-operator eigenfunction problem (\ref{RadialIntEq}) to an equivalent eigenfunction problem for a second-order differential operator with which it commutes and thus shares common eigenfunctions. Since the differential operator is closely related to that which annihilates the Zernike polynomials, the coefficients decay rapidly with increasing $k$, and the infinite dimensional TDM may be truncated at some large but finite dimension, depending on the number of eigenvalues and eigenvectors that are desired, without incurring significant error. Open-source Matlab codes have been developed and made available by Lederman \cite{Lederman16} to compute the superposition coefficients and the associated eigenvalues. 

In the present paper,  we have developed, as we show in Appendix B, an alternative approach to derive the associated eigenfunctions, the 2D radial PSWFs, $\Psi_{Nn}(r)$, in their Bessel form and computed the associated eigenvalues $\lambda_{Nn}$ by numerically diagonalizing an infinite-dimensional matrix system of equations to which the eigenvalue equation (\ref{eigenrelation2a}) is equivalent along the radial dimension.  In a rather tedious manner using certain integral identities involving the radial Zernike polynomials, Slepian \cite{Slepian64,GreengardSerkh18} derived such a Bessel form for the 2D radial PSWFs, but our approach, in sharp contrast, is simpler, more constructive, and less tedious. 

In the Bessel form (\ref{RadialPSWFBesselForm2}), 
\be
\label{RadialPSWFBesselForm}
\Psi_{Nn}(r)=\sum_{k=0}^\infty d^{(Nn)}_k {J_{N+2k+1}(cr)\over cr},
\ee
the radial PSWFs are convergent and analytic everywhere in the infinite plane, since Bessel functions are analytic over the full complex plane. The form (\ref{RadialPSWF}), to which Eq.~(\ref{RadialPSWFBesselForm})  is exactly equivalent over the unit disk, $0\leq r\leq 1$, is not convergent, however, for $r>1$. The Bessel form for the radial PSWFs will be particularly useful in investigations of 2D sources that are arbitrarily extended and not necessarily circular in shape, for which the 2D PSWFs with an appropriate choice of the SBP parameter $c$ will continue to serve as good basis functions. 

We also note that since for fixed $N$, the eigenvalues $\lambda_{Nn}$ decay rapidly to zero with increasing value of $n$, often with many orders of magnitude separating successive eigenvalues, only a small number of them need be calculated typically when $c$ is not too large. In our numerical computations, we allowed $N$ to range from 0 to a number of order $N_{\rm max}\sim c+15$. For $N=N_{max}$, the largest  eigenvalue $\lambda_{N_{\rm max}1}$ is already 8-10 orders of magnitude smaller than the largest one for values of $N$ of order 1-5.   

\subsection{QFI for Disk-Radius Estimation}
We specialize expression (\ref{Hmunu1}) for QFI for the case of a single parameter, $\mu=\nu=a$, the source-disk radius, by replacing $\hrho_1$ by $\hrho_2$ appropriate for the disk problem, combining the $i=j$ terms from the double sum in that expression with its first sum and then symmetrizing the remaining double sum over $i\neq j$. We then replace the general sum indices $i$ and $j$ each by the three indices that uniquely label each eigenstate in the range space of $\hrho_2$, namely by $N,n,p$ and $N',n',p'$, in which the indices $N, N'$ run over all non-negative integers, $n,n'$ over all positive integers, and $p,p'$, the parity indices, take the values $\pm$, so we arrive at the expression, 
\ba
\label{QFI}
&H_{aa}\!\!=\!\!\sum_{N,n,p} {1\over \lambda_{Nn}}\Big[4\langle \lambda_{Nn}^{(p)}|(\partial \hat\rho_2)^2|\lambda_{Nn}^{(p)}\rangle
-3\la\lambda_{Nn}^{(p)}|\partial\hrho_2|\lambda_{Nn}^{(p)}\ra^2\Big]\nn
&+2\sum_{(N,n,p)\neq (N',n',p')}\left[{1\over (\lambda_{Nn}+\lambda_{N'n'})}-{1\over \lambda_{Nn}}-{1\over\lambda_{N'n'}}\right]\nn
&\qquad\qquad\qquad\times|\langle \lambda_{Nn}^{(p)}|\partial\hat\rho_2|\lambda_{N'n'}^{(p')}\rangle|^2.
\end{align}
The symbol $\partial$ denotes the partial derivative of the very first quantity to its immediate right w.r.t.~$a$. For this single parameter of interest, QFI is a number equal to $H_{aa}$, with its reciprocal furnishing the lowest possible bound, QCRB, on the variance of {\em any} unbiased estimation of the source-disk radius.

In order to evaluate the two different kinds of matrix elements needed here, namely $\la \lambda_{Nn}^{(p)}|\partial\hrho_2|\lambda_{N'n'}^{(p')}\ra$ and $\la \lambda_{Nn}^{(p)}|(\partial\hrho_2)^2|\lambda_{Nn}^{(p)}\ra$, we start by taking the partial derivative of expression (\ref{srho12}) for $\hrho_2$ w.r.t.~$a$ and then squaring that derivative, which yields the expressions,
\ba
\label{prho}
&\partial\hrho_2 = {1\over \pi}\int_{\cD_1}dA\,\big [\partial |K_{a\br}\ra \la K_{a\br}|+|K_{a\br}\ra\partial\la K_{a\br}|\big],\nn
&(\partial\hrho_2)^2 = {1\over \pi^2}\int_{\cD_1}\int_{\cD_1}dA\, dA'\, \big[\partial |K_{a\br}\ra \la K_{a\br}|\partial|K_{a\brp}\ra\la K_{a\brp}|\nn
&+\partial |K_{a\br}\ra \la K_{a\br}|K_{a\brp}\ra\partial \la K_{a\brp}|
+|K_{a\br}\ra\partial\la K_{a\br}|\partial|K_{a\brp}\ra\la K_{a\brp}|\nn
&+|K_{a\br}\ra\partial\la K_{a\br}|K_{a\brp}\ra\partial\la K_{a\brp}|\big].
\end{align}
We next compute the matrix elements of these expressions by noting the inner-product identity,
\be
\label{KNn}
\la K_{a\br}|\lambda_{Nn}^{(p)}\ra = \lambda_{Nn}C_{Nn}^{(p)}(\br),\ 0\leq r\leq 1,
\ee
which follows from taking the inner product of $\la K_{a\br}|$ with the eigenvector expansion (\ref{expansion}), with $\br$ replaced by $\brp$ inside the integral, and then using the eigenrelation (\ref{eigenrelation2a}) in which we recognize $O_a(|\br-\brp|)$ as the inner product $\la K_{a\br}|K_{a\brp}\ra$.

In view of identity (\ref{KNn}) and its complex conjugate, we may reduce the matrix elements of expressions (\ref{prho}) to the form,
\ba
\label{prhoMatEl}
\la \lambda_{Nn}^{(p)}|\partial&\hrho_2|\lambda_{N'n'}^{(p')}\ra = {1\over \pi}\int_{\cD_1}dA\,\big [ \lambda_{N'n'}C_{N'n'}^{(p')}(\br)L_{Nn}^{(p)}(\br)\nn
&+\lambda_{Nn}C_{Nn}^{(p)}(\br)L_{N'n'}^{(p')}(\br) \big],\nn
\la \lambda_{Nn}^{(p)}|(\partial\hrho_2)^2&|\lambda_{Nn}^{(p)}\ra = {1\over \pi^2}\int_{\cD_1}\int_{\cD_1}dA\, dA'\, \Bigg\{ L_{(Nn)}^{(p)}(\br)\nn
\times&\Big[P_a(\br,\brp)\lambda_{Nn}C_{Nn}^{(p)}(\brp)+O_a(\br-\brp)L_{Nn}^{(p)}(\brp)\Big]\nn
+&\lambda_{Nn}^{(p)}(\br)\,C_{Nn}^{(p)}(\br)\Big[ Q_a(\br,\brp)\, \lambda_{Nn}C_{Nn}^{(p)}(\brp)\nn
+&P_a(\brp,\br)L_{Nn}^{(p)}(\brp)\Big]\Bigg\},
\end{align}
where the functions $L_{Nn}^{(p)}$, $P_a$, and $Q_a$ are defined as 
\ba
\label{LPQ}
L_{Nn}^{(p)}(\br)=&\la \lambda_{Nn}^{(p)}|\partial|K_{a\br}\ra,\ \ 
P_a(\brp,\br) =\la K_{a\brp}|\partial|K_{a\br}\ra,\nn 
Q_a(\br,\brp)=&\partial\la K_{a\br}|\partial|K_{a\brp}\ra,
\end{align}
and we have assumed, as we confirm by means of a detailed evaluation of these functions in Appendix C, that these functions are all real. The first of the matrix elements in Eq.~(\ref{prhoMatEl}) evaluates to the form,
\ba
\label{prhoMatEl1}
\la \lambda_{Nn}^{(p)}|\partial\hrho_2&|\lambda_{N'n'}^{(p')}\ra =\delta_{NN'}\delta_{pp'} {1\over \pi}\int_0^1 dr\, r\,\big [ \lambda_{Nn'}\Psi_{Nn'}(r)\nn
&\times L_{Nn}(r)+\lambda_{Nn}\Psi_{Nn}(r)L_{Nn'}(r) \big],
\end{align}
which vanishes unless $(N,p)=(N',p')$. Expressions for $\Psi_{Nn}(r)$ and $L_{Nn}(r)$, which are the radial parts of the full functions, $C_{Nn}^{(p)}(\br)$ and $L_{Nn}^{(p)}(\br)$, respectively, are also derived in Appendix C. 

Consider now the second matrix element in Eq.~(\ref{prhoMatEl}). Since the functions $C_{Nn}^{(p)}(r,\theta)$ and $L_{Nn}^{(p)}(r,\theta)$ are both separable in their radial and angular factors, while the functions $O_a (\br,\brp), P_a(\br,\brp),$ and $Q_a(\br,\brp)$ all depend on the angles through $\cos(\theta'-\theta)$, it follows that one of the two angular integrals, say the one over $\theta$, can be evaluated quite trivially after the other angle, $\theta'$, has been shifted by $\theta$. 
This set of steps, when combined with the symmetry of the double area integral on the RHS of the second of the expressions (\ref{prhoMatEl}) that renders its first and fourth terms equal, allows us to simplify it. Specifically, this yields a sum of three terms, each involving a double radial integral, as
\ba
\label{prhoMatEl2}
\la &\lambda_{Nn}^{(p)}|(\partial\hrho_2)^2|\lambda_{Nn}^{(p)}\ra = {1\over \pi^2}\int_0^1 dr\,r\, \int_0^1dr'\, r'\, \big\{
2\lambda_{Nn}\nn
&\times \Psi_{Nn}(r)P_N(r',r)\,L_{Nn}(r')
+L_{Nn}(r) O_N(r,r')\,L_{Nn}(r')\nn
& +\lambda_{Nn}^2\,\Psi_{Nn}(r)Q_N(r,r') \, \Psi_{Nn}(r')\big\},
\end{align} 
where $O_N$, $P_N$, and $Q_N$ are the following integrals over the angular difference, $\theta'-\theta$, relabeled as $\theta'$:
\ba
\label{OPQN}
O_N(r,r')=&\oint d\theta' O_a(r,r',\theta')\cos N\theta',\nn
P_N(r',r)=&\oint d\theta' P_a(r',r,\theta')\cos N\theta',\nn
Q_N(r,r')=&\oint d\theta' Q_a(r,r',\theta')\cos N\theta',
\end{align}
in which $P_a(r',r,\theta')$ has already been defined in Eq.~(\ref{Pintegral1}) as being the value of $P_a(\brp,\br)$ at angle $\theta=0$, and $O_a(r,r',\theta')$ and $Q_a(r,r',\theta')$ are similarly the values of $O_a(\br,\brp)$ and $Q_a(\br,\brp)$ at $\theta=0$, respectively, that may be obtained from Eqs.~(\ref{overlap1}) and (\ref{Qintegral1}) as
\ba
\label{OQ}
O_a(r,r',\theta') = &2{J_1(X_0)\over X_0},\nn
Q_a(r,r',\theta')=& 4\pi^2 rr'\Bigg\{c^2[2rr'-(r^2+r^{'2})\cos\theta']{J_3(X_0)\over X_0^3}\nn
&+\cos\theta'\left({J_1(X_0)\over X_0}-2{J_2(X_0)\over X_0^2}\right)\Bigg\}.
\end{align}
We also note that $L_{Nn}(r)$, given by Eq.~(\ref{rLNn}), may be expressed more simply in terms of $P_N$ as
\be
\label{rLNn1}
L_{Nn}(r) = {1\over\pi}\int_0^1 dr'\, r' \Psi_{Nn}(r')\,P_N(r',r) \cos N\theta',
\ee
while $\Psi_{Nn}$ is given by Eq.~(\ref{RadialPSWFBesselForm}) with $c=2\pi a$.

A numerical evaluation of the matrix elements (\ref{prhoMatEl1}) and (\ref{prhoMatEl2}) needed for QFI (\ref{QFI}) requires first computing the coeffiicients $h_{Nnk}$ of Eq.~(\ref{RadialPSWFBesselForm}) numerically, which we did using Lederman's codes \cite{Lederman16}, and then evaluating the angular integrals (\ref{OPQN}) before evaluating the radial integrals (\ref{rLNn1}). Since the latter matrix element involves a double radial integral of an integrand involving the angular integrals (\ref{OPQN}), its direct numerical computation is quite inefficient. We overcome this inefficiency by making use of Gegenbauer's addition theorem \cite{Watson95} for Bessel functions, 
\ba
\label{Gegenbauer}
{J_n(X_0)\over X_0^n}=&2\sum_{m=0}^\infty {J_{m+n}(cr)\over (cr)^n}{J_{m+n}(cr')\over (cr')^n}\nn
&\qquad\times {d^n\over d(\cos\theta')^n}\cos[(m+n)\theta'],\nn
&X_0=c\sqrt{r^2+r^{'2}-2rr'\cos\theta'},
\end{align}
which, in light of expressions (\ref{OPQN}) and (\ref{OQ}), allows us to express the matrix element (\ref{prhoMatEl2}) as a sum over the index $m$ of a number of factorized, separable $r$ and $r'$ integrals involving only three values of $n$, namely 1,2, and 3. The computational efficiency is greatly improved since the number of terms that contribute significantly to the sum (\ref{Gegenbauer}) is at most of order $c+15$ or so, since the Bessel functions decay super-exponentially when their order exceeds their argument.  

We can now compute expression (\ref{QFI}) for the single-photon QFI w.r.t. the source disk radius $a$ by noting that the diagonal matrix elements inside the sum in the first line of this expression are independent of the parity index $p$, while according to Eq.~(\ref{prhoMatEl1}) the off-diagonal elements vanish unless $N=N'$ and $p=p'$. As a result, in the first sum over $p$, for any given $(Nn)$ pair, in Eq.~(\ref{QFI}) simply yields a factor equal to the degeneracy of the state $|\lambda_{Nn}^{(p)}\ra$, which we denote as $\pi_N$ and which, as noted earlier, takes the values,
\be
\label{degeneracy}
\pi_N=\left\{
\begin{array}{ll}
1,& N=0\\
2, & N\geq 1,
\end{array}
\right.
\ee
while the six-fold ``off-diagonal" sum in the second line of Eq.~(\ref{QFI}) reduces to a mere triple sum over $N$, $n$, $n'$, with $N'=N$, $n\neq n'$, with the degeneracy factor, $\pi_N$, modifying the sum over $N$,
\ba
\label{QFI1}
H_{aa}=&\sum_{N=0}^\infty\pi_N\sum_{n=1}^\infty{1\over \lambda_{Nn}}\nn
            &\times\left[4\langle \lambda_{Nn}^{(p)}|(\partial \hat\rho_2)^2|\lambda_{Nn}^{(p)}\rangle-3\la\lambda_{Nn}^{(p)}|\partial\hrho_2|\lambda_{Nn}^{(p)}\ra^2\right]\nn
+&2\sum_{N=0}^\infty \pi_N\sum_{n\neq n'=1}^\infty\left[{1\over (\lambda_{Nn}+\lambda_{Nn'})}-{1\over \lambda_{Nn}}-{1\over\lambda_{Nn'}}\right]\nn
&\times|\langle \lambda_{Nn}^{(p)}|\partial\hat\rho_2|\lambda_{Nn'}^{(p)}\rangle|^2.
\end{align}

To evaluate expression (\ref{QFI1}) for QFI numerically, we first truncated the sums over $N,n,n'$ at large enough upper integral values that the smallest eigenvalues associated with the terms of the truncated sums were all larger than a certain threshold, which we chose as $10^{-9}$. Because of the rapid decay of the eigenvalues $\lambda_{Nn}$ with increasing values of $N$ and $n$, the neglected terms in these sums contribute negligibly, as we checked numerically, validating the truncation \cite{footnote1}. 
But before we display our numerical results for QFI, we calculate CFI w.r.t.~the classical Zernike projection basis for estimating the disk radius.

\subsection{Radius Estimation Using Zernike Wavefront Projections}
Consider wavefront projection measurements of disk emission in the pupil plane in the Zernike basis and the probability of the photon wavefront being measured in each of the first four Zernike modes defined in Eq.~(\ref{Z1234}). From these probabilities, we will evaluate the classical Fisher information (CFI) for estimating the disk radius. 

The mode projection probabilities are defined as the squared moduli of the overlap integrals between the normalized photon wavefunction and the individual Zernike modes, averaged over the incoherently but uniformly illuminated source disk,
\be
\label{ZProbs}
 P_j=\la Z_j|\hrho_2|Z_j\ra, \ j=0,1,2, \ldots.
 \ee
In view of expression (\ref{srho12}) for the photon SPDO $\hrho_2$ defined in a conveniently scaled form and the wavefunction (\ref{wavefunction}) for point-source emission,  we may arrive at the integral form,
\be
\label{ZProbs1}
P_j={1\over \pi^2 }\int \!d^2r\, P(\br) \left\vert\int\! d^2u\,P(\bu)\,\exp(-i2\pi a\br\cdot\bu) Z_j(\bu)\right\vert^2,
\ee
for these probabilities. Since expression (\ref{ZProbs1}), apart from the outer integral over the unit disk, is formally the same as the corresponding expression (\ref{ProbZj}) for the line-source problem discussed in Sec.~III.C, we may evaluate the former in an entirely analogous manner, as we show in Appendix A. The final expression for $F_{aa}^{(\infty)}$ takes the form,
\ba
\label{Faa_full}
F_{aa}^{(\infty)} =& {4\over a^2}\Bigg[{1\over 2\pi^2 a^2}\sum_{p=0}^\infty(p+1)^2{J_{p+1}^4(2\pi a)\over \int_0^{2\pi a} dw \,J_{p+1}^2(w)/w}-1\Bigg].
\end{align}

In view of the small-argument expansion, $J_{p+1}(w)\sim w^{p+1}$ for  $w<<1$, $P_{pm\sigma}$ given by Eq.~(\ref{ZProbs}) is of order $O(a^{2p})$. As such, only the two partial contributions, $F_{aa}^{(11\pm)}$, to the total CFI, as seen by the form of Eq.~(\ref{Faa_pms}), are non-vanishing in the limit of highly sub-diffractive radius, $a<<1$. Other partial contributions to $F_{aa}$ tend, however, to bring it closer to QFI over non-zero values of $a$. 

\subsection{Numerical Results}

In Fig.~4, we display the numerically calculated values of QFI per photon emitted by a disk of radius $a$ and processed by the imaging system as a function of $a$, which is the parameter to be estimated. The decrease of QFI with increasing disk radius, as shown by the solid curve with open circles at the computed points, is expected since as the radius grows, the photons emitted from anywhere on the disk contain increasing less information about how large the disk is, with photons emitted at the disk boundary being the only ones carrying such information. Since the emission probability of a photon is uniform over the disk, the probability for its emission from within a diffraction width, of order 1 in scaled units, of the disk perimeter is only $2\pi a\times 1/\pi a^2\sim 1/a$ whenever $a>>1$. This implies a reduction of the radius-estimation fidelity, or QFI, according to an inverse linear dependence on the radius $a$, which is well verified in the dashed-curve fit on the figure for values of $a$ greater than about 1.5.

On the figure, we also plot the dependence of CFI, $F_{aa}^{(N)}$, for estimating the disk radius by data obtained when only a small number, $N$, of the lowest-order Zernike wavefront projections are measured, for which CFI is given by Eq.~(\ref{Fmunu}) for the special case, $\mu=\nu=a$, 
\be
\label{FaaN}
F_{aa}^{(N)}= \sum_{j=1}^N {1\over P_j}\left({\partial P_j\over \partial a}\right)^2+{1\over \bar P}\left({\partial \bar P\over \partial a}\right)^2,
\ee
with $\bar P=1-\sum_{j=1}^N P_j$ being the probability of the {\em unobserved} mode projections and $P_j$ given by expression (\ref{ZProbs}). Specifically, we plot $F_{aa}^{(N)}$ when either the tip and tilt ($Z_2,Z_3$) mode projections, or the tip, tilt, and piston ($Z_1,Z_2, Z_3$) mode projections, or the tip, tilt, piston, and defocus ($Z_1,\ldots,Z_4$) mode projections are the only ones that are measured. We see that mode projections in the tip-tilt modes alone can attain QFI in the extreme superresolution limit of vanishing radius. Adding piston projection data to the tip-tilt projection data broadens the CFI plot without altering its peak value, indicating that further enhancements of the estimation fidelity of the wavefront projection approach result for finite but still subdiffractive radius values when additional low-order Zernike mode projections are included. The addition of the defocus mode, $Z_4$, yields a rather dramatic enhancement of CFI, bringing it quite close to QFI, for values of $a$ between 0.1 and 0.5 or so. 

We expect the improvement to continue and the gap between QFI and CFI to shrink ever more for still larger values of $a$ as more of the higher-order Zernike modes are added to the set of observed projections. But, as we can see from the plot of the full CFI, $F_{aa}^{(\infty)}$, on including all Zernike mode projections, the QFI-CFI gap is, in fact, not fully bridged even in this case. As we noted earlier for the case of estimating the line-source parameters, the persistence of this gap could indicate that QFI is in fact unattainable, with its inverse providing only a loose lower bound on the variance of {\em any} unbiased estimator of the disk radius. It is also possible that the Zernikes may not constitute the most optimal basis for estimating the radius via wavefront projections.
\begin{figure}
\centerline{
\includegraphics[width=0.5\textwidth]{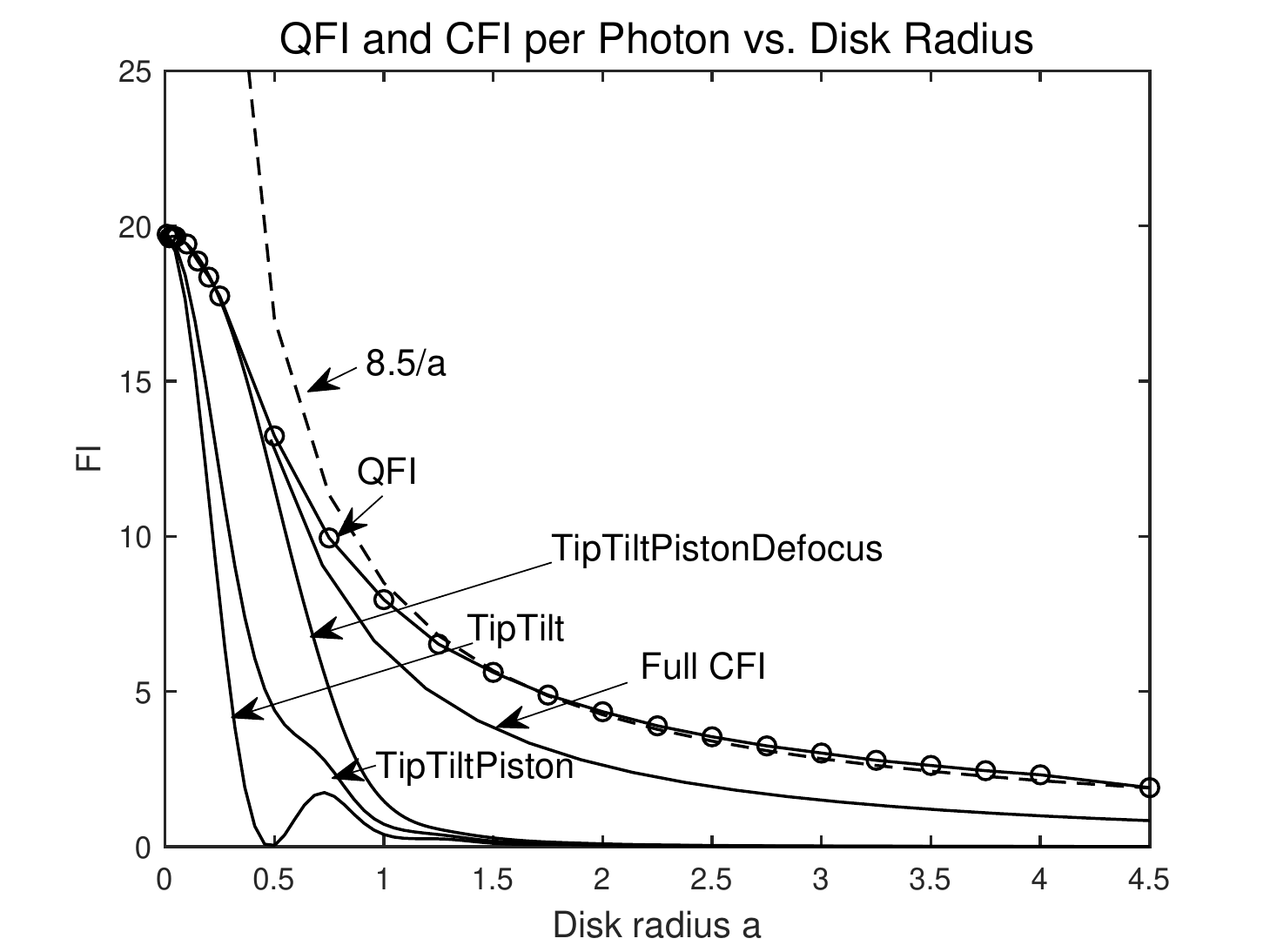}}
\caption{QFI and CFI for estimating the disk radius $a$. The dashed curve is an asymptotic fit to QFI for large values of $a$, while the three lowest solid curves are plots of CFI for tip-tilt, tip-tilt-piston, and tip-tilt-piston-defocus Zernike projections, respectively.}
\end{figure}

\section{QFI for a Uniformly Bright Source of Arbitrary Parametrizable Shape}

We now generalize our QFI calculations to a finite 2D source of uniform brightness bounded by a general, parametrizable curve that is specified in polar coordinates $(r,\theta)$ as $r=f(\theta;\bsigma)$, where $\bm{\sigma}\defeq(\sigma_1,\ldots,\sigma_P)$ is the vector of $P$ parameters that control the spatial extension and orientation of the source. We assume that the boundary shape is known {\em a priori}, and $f(\theta;\bm{\sigma})$ is a well behaved, single-valued function of $\theta$ and its parameter vector $\bm{\sigma}$, the latter of which must be estimated from image data. 

If $\cA$ is the area of the source support, then $a=\sqrt{\cA/\pi}$, which we call the radius parameter of the source, characterizes its linear extension about its center. The SPDO for this source may be written as
\be
\label{Brho}
\hrho={1\over \pi}\int_{\cD_a}dA\, |K_{a\br}\ra\la K_{a\br}|,
\ee
in which $\cD_a$ denotes the interior of the planar 2D source after the isotropic scaling, $\br\to a\br$, with the scaling factor $1/a$. The scaled source has, by construction, area equal to that of the unit disk, namely $\pi$. 

An eigenstate $|\lambda\ra$ of $\hrho$ with eigenvalue $\lambda$ may be expanded in the same way over the scaled source area,
\be
\label{Bexpansion}
|\lambda\ra = {1\over \pi}\int_{\cD_a} dA\,C_\lambda(\br)|K_{a\br}\ra.
\ee
Substituting expressions (\ref{Bexpansion}) and (\ref{Brho}) into the eigenrelation, $\hrho|\lambda\ra = \lambda|\lambda\ra$, yields the integral equation obeyed by the coefficient function $C_\lambda(\br)$ as
\be
\label{Bint_eq}
{1\over \pi}\int_{\cD_a}O_a(|\br-\br')\, \Cl(\br')\,dA' = \lambda \Cl(\br), \ \br\in \cD_a,
\ee
in which $O_a(|\br-\br'|)$ is given by Eq.~(\ref{overlap}). Since the kernel function $O_a(|\br-\brp|)$ admits the Gegenbauer decomposition (\ref{BesselAddThm}) in terms of radial Bessel functions, the radial dependence of $\Cl(r,\theta)$ may also be expanded in terms of such Bessel functions, while its angular dependence can be expanded in the compler Fourier series as
\be
\label{Bcoeff_expansion}
\Cl(\br)=\sum_{m=0}^\infty\sum_{n=-\infty}^\infty (m+1)^{1\over 2}C_{mn} {J_{m+1}(2\pi a r)\over 2\pi a r}\exp(in\theta),
\ee
in which an extra factor of $(m+1)^{1/2}$ has been included for later convenience. A substitution of expansion (\ref{Bcoeff_expansion}) along with the Gegenbauer addition formula (\ref{BesselAddThm})   into the integral equation (\ref{Bint_eq}), followed by equating the coefficients of $J_{m+1}(2\pi ar)$ on both sides, generates the following equation valid for each value of $m$:
\ba
\label{Bint_eq_theta}
{4\over \pi}&\sum_{m',n'}C_{m'n'}[(m+1)(m'+1)]^{1\over 2}\int_{\cD_a} dA' {J_{m+1}(2\pi a r')\over 2\pi a r'}
\nn
&\times {J_{m'+1}(2\pi a r')\over 2\pi a r'}{\sin(m+1)(\theta-\theta')\over \sin(\theta-\theta')} \exp(in'\theta')\nn
&=\lambda\sum_n C_{mn}\exp(in\theta).
\end{align}
Multiplying both sides of Eq.~(\ref{Bint_eq_theta}) by $\exp(-ip\theta)$ and integrating over the full $2\pi$ period of $\theta$, we may extract the individual coefficient $C_{mp}$ on the RHS of this equation. Relabeling the $p$ index as $n$, we thus obtain the following reduced equation: 
\ba
\label{BCmn}
{4\over \pi}&A_{mn}\sum_{m',n'}C_{m'n'}[(m+1)(m'+1)]^{1\over 2}\int_{\cD_a} dA' {J_{m+1}(2\pi a r')\over 2\pi a r'}\nn
&\times {J_{m'+1}(2\pi a r')\over 2\pi a r'}\exp[-i(n-n')\theta']=\lambda C_{mn},
\end{align}
where $A_{mn}$ is defined as the integral,
\be
\label{BAmn}
A_{mn}={1\over 2\pi}\oint d\theta {\sin(m+1)\theta\over \sin\theta}\exp(-in\theta).
\ee
To simplify the integral over $\theta$, we first performed the shift, $\theta\to\theta+\theta'$,  which does not alter its range of integration, that being the full $2\pi$ period of the integrand, and then combined the extra $\exp(-in\theta')$ factor that resulted with the factor $\exp(in'\theta')$ present inside the integral over $dA'$ to arrive at expression (\ref{BAmn}) for $A_{mn}$.

By noting identity (\ref{trig2}) in its complex form,
\be
\label{trig2b}
{\sin (m+1)\theta\over \sin\theta}= \sum_{k=-m,-m+2,\ldots}^m \exp(ik\theta),
\ee  
and substituting it into the integral (\ref{BAmn}), we can easily evaluate $A_{mn}$ as
\be
\label{BAmn1}
A_{mn}=\left\{
\begin{array}{ll}
1, & m-|n|=0,2,\ldots\\
0, & {\rm otherwise}
\end{array}
\right.
\ee
Due to result (\ref{BAmn1}), the allowed ranges of the coefficient index pair, $m,n$, are $m=0,1,\ldots;\ n=-m,-m+2,\ldots, m$, so the double sum in expansion (\ref{Bcoeff_expansion}), for each value of index $m$, is restricted in the index $n$ as just stated, 
\be
\label{Bcoeff_expansion1}
\Cl(\br)=\sum_{m,n} (m+1)^{1\over 2}C_{mn} {J_{m+1}(2\pi a r)\over 2\pi a r}\exp(in\theta),
\ee
and the double sum in Eq.~(\ref{BCmn}) too reduces similarly,
\ba
\label{BCmn1}
\sum_{m',n'} S_{mn;m'n'}C_{m'n'}&=\lambda C_{mn},\ m=0,1,\ldots,\nn
                  &\quad n=-m,-m+2,\ldots,m,
\end{align}
in which the system array elements $S_{mn;m'n'}$ are defined as
\ba
\label{BSmnmn}
S_{mn;m'n'}=&{4\over \pi}[(m+1)(m'+1)]^{1\over 2}\int_{\cD_a} dA' {J_{m+1}(2\pi a r')\over 2\pi a r'}\nn
&\times {J_{m'+1}(2\pi a r')\over 2\pi a r'}\exp[-i(n-n')\theta'].
\end{align}
We immediately note that array $S$ is Hermitian in its two pairs of indices, namely
\be
\label{BMhermitian}
S^*_{mn;m'n'}=S_{m'n';mn}.
\ee
Any double sum, like those in Eqs.~(\ref{Bcoeff_expansion1}) and (\ref{BCmn1}), will henceforth be understood, unless explicitly indicated otherwise, to be defined with its first, outer-sum index, $m$, running over all non-negative integers and the second index, $n$, running from $-m$ to $m$ in steps of 2. 

Equation (\ref{BCmn1}) may be regarded as a matrix equation if we map the pair of indices $(m,n)$ onto a single integer index that counts the various coefficients, $C_{mn}$, starting with $C_{00}$. Since the system matrix is Hermitian and non-negative definite, the eigenvalues are all real and non-negative. Furthermore, they add up to 1, since the sum of the diagonal elements, $S_{mn;mn}$ over all allowed values of $m$ and the $(m+1)$ values of $n$, for each value of $m$, can be easily shown to be 1 using the Gegenbauer sum rule (\ref{sumrule}) and the fact that the scaled source area is $\pi$, as we noted earlier. The matrix elements also decay super-exponentially with order $m$, whenever $2\pi a$ is not too large compared to 1, since $J_{m+1}(x)/x\to (x/2)^m/ m!$ for $|x|^2<<m$. This implies, as we have already noted for the line and disk source problems, that the matrix can be truncated at some finite upper value of $m$, say $M$, that is only large compared to the square of the effective SBP, $2\pi a$. Since for each value of $m$, there are $(m+1)$ terms in the $n$ sum, the system matrix, $\bS$, has dimension $M_s\times M_s$, where
\be
\label{Ms}
M_s=\sum_{m=0}^M (m+1) ={(M+1)(M+2)\over 2}.
\ee

The quadratic scaling of the linear dimension of the system matrix with the upper cutoff of index $m$ is equivalent to a quartic scaling in the characteristic size, $a$, of the source when $a$ is large. This implies that the number of optimal wavefront projections needed to achieve the highest possible resolution allowed by QFI must also scale quartically with the source size in the large-size limit. The conventional intensity based imaging, on  the other hand, may seem to achieve this with only a quadratic scaling in the linear dimensions of the source according to which the number of pixels in the image scale. However, to reach sub-diffractive scales of resolution in local regions of the image, the photon cost for conventional imaging can be prohibitively large at its inverse quartic scaling with the sought resolution scale when compared to the wavefront projection based approach that suffers from only an inverse quadratic photon cost  for such local superresolution imaging. These trade-offs between the photon cost and number of modal projections needed w.r.t.~the operating requirements for superresolution in a scene might inform a hybrid approach that applies the wavefront projection based superresolution imaging in some areas of the image and conventional intensity based imaging in other image areas.

Numerically evaluating the area integrals in Eq.~(\ref{BSmnmn}) for a number of $m,n$ values may still be tedious and prohibitive. The use of indefinite-integral identities (\ref{BesselProdInt1}) and (\ref{BesselInt5}), however, can help reduce these area integrals to simple angular integrals that can be computed efficiently. We see this by writing $dA'$ as $d\theta'\, r'\,dr'$ and recognizing that for a given $\theta'$, the limits on the $r'$ integral are 0 and $f(\theta';\bsigma)/a$, so the area integral (\ref{BSmnmn}) reduces to a simple angular integral, 
\ba
\label{BAreaInt}
&\int_{\cD_a} dA' {J_{m+1}(2\pi a r')\over 2\pi a r'} {J_{m'+1}(2\pi a r')\over 2\pi a r'}\exp[-i(n-n')\theta']\nn
&={1\over 4\pi\cA}\oint d\theta'\exp[-i(n-n')\theta']\, I_{mm'}(\theta';\bsigma),
\end{align}
in which $\pi a^2$ was replaced by the source area, $\cA$, and $I_{mm'}$, defined as
\be
\label{BImm}
I_{mm'}(\theta';\bsigma) = \int_0^{2\pi f(\theta';\bsigma)} {J_{m+1}(x) J_{m'+1}(x)\over x} dx,
\ee
where $x=2\pi ar'$, has already been evaluated in closed form in Eqs.~(\ref{BesselProdInt1}) and (\ref{BesselInt5}). Further, being of the Fourier form, the angular integral (\ref{BAreaInt}) may be efficiently evaluated by the fast Fourier transform (FFT).

\subsection{Computation of QFI}

In addition to the eigenvalues and eigenvectors of SPDO (\ref{Brho}) that we can calculate numerically via the matrix approach just outlined, we must also compute the first partial derivatives of $\hrho$ in order to use formula (\ref{Hmunu1}) for QFI. To do so, we first write expression (\ref{Brho}) for SPDO in an equivalent form using the indicator function, $\Theta_\cD(\br)$,  for the source area in terms of the original unscaled position vector $\br$,
\be
\label{Brho1}
\hrho={1\over \cA}\int dA\, \Theta_\cD(\br)\, |K_\br\ra\la K_\br|,
\ee
where the area integral is now formally over all space. Since the source occupies a singly connected domain, $\cD$, with its boundary curve, $\partial D$, specified by a single-valued function, $r=f(\theta;\bm{\sigma})$, the indicator function $\Theta_\cD(\br)$ for the source may then be written in terms of the unit step function as $\Theta(f(\theta;\bsigma)-r)$, and Eq.~(\ref{Brho1}) transforms formally to the unbounded integral,
\be
\label{Brho2}
\hrho={1\over \cA}\int dA\, \Theta(f(\theta;\bsigma)-r)\, |K_\br\ra\la K_\br|.
\ee
The partial derivative of $\hrho$ w.r.t.~parameter $\sigma_\mu$ may now be calculated as
\ba
\label{Bpmu_rho}
\pmu&\hrho={1\over \cA}\Big[-\hrho\,\pmu\cA \nn&+\int dA\, \delta(r-f(\theta;\bsigma))\,\pmu f(\theta;\bsigma)
|K_\br\ra\la K_\br|\Big]\nn
&\quad ={1\over \cA}\Big[-\hrho\,\pmu\cA \nn&+\oint d\theta\, f(\theta;\bsigma)\,\pmu f(\theta;\bsigma)
|K_{f(\theta,\bsigma),\theta}\ra\la K_{f(\theta,\bsigma),\theta}|\Big],
\end{align}
in which we used the identity, 
\be
\label{Bdelta}
{d\over dx}\Theta(a-x) =-\delta(x-a),
\ee
to differentiate the step function inside the integral.

Note that all off-diagonal elements of the first term on the RHS of Eq.~(\ref{Bpmu_rho}) in the SPDO eigenbasis vanish. The second term there may be calculated quite simply in terms of the coefficient functions using the identity,
\be
\label{Bidentity}
\la K_\br|\lambda_i\ra = \lambda_i C_i(\br/a),
\ee
which is obtained, like similar previous relations, by substituting expression (\ref{Bexpansion}) on the LHS and then using the integral equation (\ref{Bint_eq}), in which a return to the original unscaled spatial position vector $\br$ has been effected by the transformation $a\br \to \br$. The resulting expression for the matrix elements of $\pmu\hrho$ is thus of form,
\ba
\label{Bdrhoij}
\la \lambda_i|&\pmu\hrho|\lambda_j\ra = -\delta_{ij} {\lambda_i\over \cA} \pmu \cA\nn
                     &+{\lambda_i\lambda_j\over\cA}\oint d\theta\, f\pmu fC_i^*(f/a,\theta)\, C_j(f/a,\theta),
\end{align}
in which we have suppressed the arguments, $\theta,\bsigma$, from the function $f$ for brevity of notation. 

On multiplying $\pmu\hrho$ given by Eq.~(\ref{Bpmu_rho}) by a similar expression for $\pnu\hrho$ given by replacing  $\mu$ by $\nu$ and $\theta$ by $\theta'$ in that equation and then constructing the diagonal matrix elements of the product, we obtain
\ba
\label{Bdrhodrhoii}
\la\lambda_i|&\pmu\hrho\,\pnu\hrho|\lambda_i\ra = {\lambda_i^2\over\cA^2}\Bigg[\pmu\cA\, \pnu\cA
         +\oint\oint d\theta d\theta' ff'\pmu f\pnu f'\nn
         &\times O_a(|\br(\theta)-\br(\theta')|/a) C_i^*(\br(\theta)/a)\,C_i(\br(\theta')/a)\nn
         &-\lambda_i\oint d\theta f\big(\pmu\cA\pnu f+\pnu\cA\pmu f\big)\,|C_i(\br(\theta)/a)|^2\Bigg],
\end{align} 
in which $\br(\theta)$ is the position vector of a point on the boundary at polar angle $\theta$ and $f=f(\theta;\bsigma)$, $f'=f(\theta';\bsigma)$. With expressions (\ref{Bdrhoij}) and (\ref{Bdrhodrhoii}) for the matrix elements of the SPDO derivatives and their bilinear products in hand, we may now evaluate QFI according to Eq.~(\ref{Hmunu1}) in terms of the coefficient functions $C_i(\br(\theta)/a)$ and simple and double integrals over the boundary of the source involving those functions and the radial coordinate function of the boundary, namely $f(\theta;\bsigma)$. 

The double integral in Eq.~(\ref{Bdrhodrhoii}) can be converted, by use of the Gegenbauer addition theorem (\ref{BesselAddThm}) along with identity (\ref{trig2b}), into a double sum over $m$ and $k$ of double integrals over $\theta$ and $\theta'$ of products of functions of $\theta$ and $\theta'$ that are complex conjugates of each other,
\ba
\label{Bdoubleintegral}
&\oint\oint d\theta d\theta' ff'\pmu f\pnu f'\,O_a(|\br(\theta)-\br(\theta')|/a)\nn
&\qquad\qquad\times C_i^*(\br(\theta)/a)\,C_i(\br(\theta')/a)\nn  
&=\sum_{m,k}\oint\oint d\theta d\theta' F_{\mu;mk}^*(\theta,\bsigma)\, F_{\nu;mk}(\theta',\bsigma)\nn
&=\sum_{m,k}\left[\oint d\theta F_{\mu;mk}^*(\theta,\bsigma)\right]\left[\oint d\theta F_{\nu;mk}(\theta,\bsigma)\right]
\end{align}
where the function $F_{\mu;mk}(\theta,\bsigma)$ has the expression,
\ba
\label{BFmk}
F_{\mu;mk}(\theta,\bsigma) =&2 \sqrt{m+1}\, r\,\pmu r{J_{m+1}(2\pi r)\over 2\pi r}C_i(\br(\theta)/ a)\nn
                                 &\times \exp(ik\theta),\ \ \ \  r=f(\theta;\bsigma),
\end{align}
and, as before, the sum over $m$ runs between 0 and $\infty$, while that over $k$ runs in steps of 2 from $-m$ to $+m$. In view of the fact that Bessel functions decay super-exponentially with increasing order, we may be able to evaluate the double sum accurately by truncating it at a relatively small upper cutoff for the $m$ sum, say at $M$, for a total of $(M+1)(M+2)/2$ terms for the double sum. Furthermore, each angular integral in Eq.~(\ref{Bdoubleintegral}) is of the Fourier form and thus amenable to efficient FFT based evaluation. Since the final terms in Eqs.~(\ref{Bdrhoij}) and (\ref{Bdrhodrhoii}) too are both simple angular integrals, expression (\ref{Hmunu1}) for QFI w.r.t.~the spatial parameters for a uniformly bright source of arbitrary geometry consists of only simple angular integrals, and can thus be evaluated highly efficiently. 

\subsection{A Centered, Fixed-Orientation Elliptical Source}
As an illustration of the approach, let us consider a uniformly bright elliptical disk centered at the origin of the coordinate system and with principal axes of half lengths $a_1$ and $a_2$ that are aligned with the coordinate $x,y$ axes. Its boundary is specified in polar coordinates as
\be
\label{Bf_ellipse}
r=\left({\cos^2\theta\over a_1^2}+{\sin^2\theta\over a_2^2}\right)^{-1/2}\!\!\equiv f(\theta; a_1,a_2).
\ee
Its area has the value, $\cA=\pi a_1 a_2$, with its radius parameter being $a=\sqrt{a_1a_2}$. The parameter $a_1$ labels its semi-minor axis length, with its semi-major axis length, $a_2$, related to $a_1$ via the ellipse eccentricity, $\epsilon$, as
\be
\label{Ba2}
a_2=a_1/\sqrt{1-\epsilon^2}.
\ee
The first-order partial derivatives of $f(\theta;a_1,a_2)$, given by Eq.~(\ref{Bf_ellipse}), w.r.t.~$a_1$ and $a_2$ are easily obtained, and all the integrals in Eqs.~(\ref{Bdrhoij}) and (\ref{Bdrhodrhoii}) that determine the matrix elements involved in expression (\ref{Hmunu1}) for QFI can be numerically computed quite efficiently.  

The $2\times 2$ QFI matrix (QFIM) w.r.t.~the two semi-axis length parameters has three independent elements. Its two diagonal elements, $H_{11}$ and $H_{22}$, are sometimes known simply as QFI \cite{Liu20} w.r.t.~the two parameters, while its off-diagonal elements, $H_{12}=H_{21}$, determine fundamentally the least possible degree of mutual interference of the two parameters, with each serving as a nuisance variable w.r.t.~the estimation of the other.  
\begin{figure}
\centerline{
\includegraphics[width=0.5\textwidth]{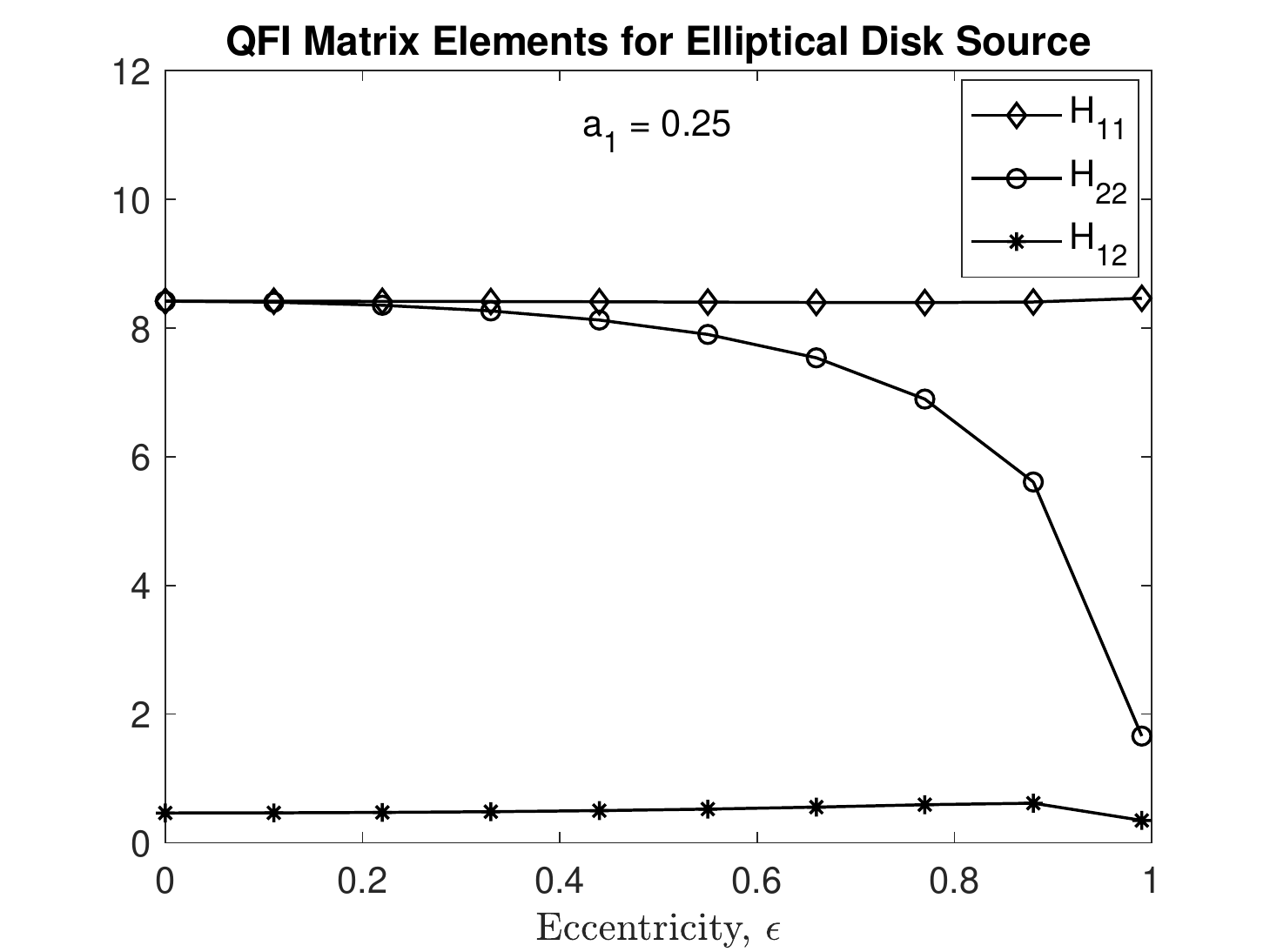}}
\caption{Plots of the three QFIM elements for estimating the semi-axis lengths of an elliptical disk vs.~its eccentricity, for $a_1=0.25$ and $a_2$ changing its value according to Eq.~(\ref{Ba2}).}
\end{figure}

In Fig.~5, we plot the three independent QFIM elements as functions of $\epsilon$, keeping the value of $a_1$ fixed at 0.25 (in units of the characteristic Airy diffraction scale defined in Sec.~II). For a number of pairs of values of $a_1$ and $a_2$ for which the QFIM elements are displayed in this plot, we checked that their final values remained essentially unchanged as we lowered the SPDO eigenvalue threshold from 10$^{-8}$ to 10$^{-12}$ for our QFI calculations via Eq.~(\ref{Hmunu1}) with its sums truncated to exclude all eigenvalues smaller than the threshold. Specifically, we observed no change of the QFIM element values out to 9 significant digits between the lower eigenvalue threshold values of $10^{-10}$ and $10^{-12}$. This observation strongly validates the excellent robustness and computational efficiency of our Bessel Fourier approach. 

For $\epsilon=0$, for which the ellipse becomes a circle, the two diagonal elements are identical, as expected.  As $\epsilon$ increases and the major axis of the ellipse lengthens relative to its minor axis, the information about the length of the major axis decreases, since an imaging photon, equally likely to be emitted from anywhere on the entire source disk, on average carries increasingly less information about the source extension. For the largest value of $\epsilon$ at 0.99 plotted in the figure, for which $a_2$ is more than 7 times larger than $a_1$, the information about $a_2$ is reduced dramatically, while information about $a_1$ remains essentially unchanged. 

Through most of the range of eccentricity values, the intereference term, $H_{12}$, increases by about 33\% from about 0.46 to 0.61, before dipping sharply for highly eccentric elliptical disks. If the estimations of the two length parameters could be made without any mutual interference, they would have contributed, in the degenerate circular limit, $\epsilon=0$, for which $a_1=a_2=0.25$, the maximum possible information about the circular radius, equal to double the information provided by an independent estimation of each parameter. That the common value, $H_{11}=H_{22}=8.45$, of QFI in this degenerate limit is slightly less than half of QFI for radius estimation, the latter being about 17.7 that we can infer from Fig.~4 at $a=0.25$, reflects the finite penalty of simultaneous estimation of the two lengths even in the degenerate limit.

\subsection{QFI for Sources of Nonuniform Brightness Distribution}

For a 2D source of an arbitrary brightness distribution, $I(\br)$, normalized to have unit area over the plane,
\be
\label{normI}
\int dA\, I(\br) = 1,
\ee
the area integrals in Eqs.~(\ref{Brho}) and (\ref{Bexpansion}) must be extended to the infinite plane and the integrand weighted by the factor $I(\br)$. These changes yield the following modified version of the integral equation (\ref{Bint_eq}) obeyed by the coefficient function $C_\lambda(\br)$:
\be
\label{Bint_eq2}
\int O_1(|\br-\br')\, I(\br')\, \Cl(\br')\,dA' = \lambda \Cl(\br),
\ee
which once again admits solutions of the complex Bessel-Fourier form (\ref{Bcoeff_expansion}). 

Correspondingly, in Eq.~(\ref{Brho1}) the factor $(1/\cA)\, \Theta_D(\br)$ must be replaced by the factor $I(\br)$ inside the integrand. Although the subsequent calculations of the SPDO eigenstates and QFI are more involved and would require more tedious numerical evaluation in the most general cases, the approach remains essentially unchanged and applicable, at least  in principle.

\section{Concluding Remarks}
In this paper we developed a formalism based on the PSWFs and their 2D generalized versions to calculate the eigenstates of SPDO and from them the QFI for estimating the spatial parameters of two simple sources located in the plane of best focus of an imager with a clear circular pupil. We then compared the numerically calculated values of QFI with the corresponding Zernike-mode-projection based CFI to assess how efficiently such wavefront projections can approach the fundamental, quantum-limited lower bounds on the theoretically attainable variance of these estimations. We calculated the two Fisher information measures specifically for estimating the length and orientation of a uniformly bright line source and the radius of a uniformly illuminated disk shaped source, both sources with fixed centers. Asymptotically the SBP associated with the PSWFs furnishes a useful measure of the number of independent degrees of freedom of the problem and thus of the characteristic dimensionality of the Hilbert space of the associated SPDO. For each source, SBP is expressed simply in terms of the ratio of the spatial dimension of the source and the Airy diffraction width mapped to the source space.
   
The PSWF based approach is readily extended to the problem of calculating the SPDO eigenfunctions and QFI for a centered circularly symmetric incoherent source with a brightness distribution that has an arbitrary radial dependence. The eigenfunctions in this case are still separable into their angular and radial factors, with the latter being conveniently expressible as superpositions of radial Bessel functions of form (\ref{RadialPSWFBesselForm}). The associated system matrix for the radial eigenfunctions will have elements that may be expressed as integrals of products of Bessel functions, much as in Eq.~(\ref{MatEl2}), with the only modification being the presence of a weight function representing the radial variation of the source intensity that multipies such products. We can thus numerically compute the eigenfunctions and eigenvalues for SPDO for such a circularly symmetric source brightness distribution (SBD). 

SPDO eigenfunctions for an extended source of still more general, non-symmetrical shape and nonuniform SBD, if its centroid is known and fixed {\em a priori}, may be expressed as linear superpositions of the full PSWFs, the latter expressed as Bessel Fourier functions of the polar coordinates.  We developed a complete formalism for computing the SPDO eigenfunctions and QFI for parameterizable sources of arbitrary shape and SBD using such Bessel Fourier basis functions. The rapid, super-exponential attenuation of the Bessel function of a fixed maximum argument with increasing order confers on the PSWF and the closely related Bessel Fourier approaches a distinct computational efficiency when compared to any purely Fourier based approach, especially when SBD has a finite support. We demonstrated high computational efficiency and accuracy of the Bessel Fourier approach for computing QFI w.r.t.~the lengths of the principal axes of a uniformly bright elliptical disk shaped source over a wide range of values for its eccentricity parameter.   

 \acknowledgments
The author is grateful for the research facilities provided by the School of Physics and Astronomy at the U. of Minnesota where he has held the position of Visiting Professor for the last two years. This work was partially supported by Hennepin Healthcare Research Institute under a research investigator appointment.
  
\appendix
\section{Zernike Based CFI}
\paragraph{The Line-Source Problem}
Using the Fourier-transform (FT) relation \cite{Noll76} obeyed by the Zernikes,
\ba
\label{ZFT}
\int &d^2u P(\bu)\exp(i2\pi x\bu\cdot\bl) \, Z_j(\bu) =\sqrt{p+1\over\pi}{J_{p+1}(2\pi xl)\over xl}\nn
&\times\left\{
\begin{array}{ll}
(-1)^{(p-m)/2}i^m\sqrt{2}\cos m\phi, & {\rm even}\ j, \, m\neq 0\\
(-1)^{(p-m)/2}i^m\sqrt{2}\sin m\phi, & {\rm odd}\ j, \, m\neq 0\\
(-1)^{p/2},& m=0,
\end{array}
\right.
\end{align}
in which $p$ and $m$, with $p\geq m$ and $p-m$ even, are the radial and azimuthal quantum numbers associated with the Zernike polynomial $Z_j$, we may express the projection probabilities for the first four Zernike modes as
\ba
\label{ProbZ1234}
P_1(\bl)=&{4\over \pi l}\int_0^{\pi l}dw{J_1^2(w)\over w^2};\nn
P_2(\bl)=&{16\over \pi l}\cos^2\phi\int_0^{\pi l}dw{J_2^2(w)\over w^2};\nn
P_3(\bl)=&{16\over \pi l}\sin^2\phi\int_0^{\pi l}dw{J_2^2(w)\over w^2};\nn
P_4(\bl)=&{12\over \pi l}\int_0^{\pi l}dw{J_3^2(w)\over w^2},
\end{align}
in which $w=2\pi l x$ serves as the new integration variable and the evenness of the integrand of the $w$ integral was used to simplify it in each case. 

By means of the identities \cite{GR96}, $J_3(w)=4J_2/r-J_1(r)$ and $J_2(w)=2J_1(w)/w-J_0(w)$ , we may turn the integrals in Eq.~(\ref{ProbZ1234}) into a sum of integrals of $J_0^2/w^4$,  $J_0(w)J_1(w)/w^3$, $J_0(w)J_1(w)/w^5$, $J_1^2(w)/w^2$, $J_1^2(w)/w^4$, and $J_1^2(w)/w^6$. The closed-form expressions for the indefinite forms of all of these integrals have been tabulated in Ref.~\cite{besint19}. We may also easily evaluate the derivatives of $P_1,\ldots,P_4$ w.r.t. $l$, as their integral expressions (\ref{ProbZ1234}) involve $l$ only in the overall coefficients multiplying the integrals and in the upper integration limit. The details of these evaluations are closely parallel to those presented in Ref.~\cite{Prasad20b}.

By including further Zernike modes beyond the first four into our projection data, we may improve CFI continually and push it closer to QFI. Consider the probability for the photon wavefront to be observed in an arbitrary Zernike mode, $Z_{pm\sigma}$, that we now label more explicitly by its radial, azimuthal, and parity indices, $p,m,\sigma$, respectively, with $\sigma=+$ and $\sigma=-$ corresponding to the $\cos m\phi$ and $\sin m\phi$ angular dependences, as defined by the FT relation (\ref{ZFT}). Use of Eq.~(\ref{ZFT}) in expression (\ref{ProbZj}) for the corresponding mode-projection probability reduces the latter to the simple form,
\ba
\label{ProbZpms}
P_{pm\sigma}=&{4(p+1)\over \pi l}\int_0^{\pi l} dw {J_{p+1}^2(w)\over w^2}\Big\{2\delta_{\sigma,-}\sin^2 m\phi\nn
&+ \delta_{\sigma,+}\left[\delta_{m0}+2\left(1-\delta_{m0}\right)\cos^2 m\phi\right]\Big\}.
\end{align}
The complicated expression inside the braces in Eq.~(\ref{ProbZpms}) is a single-line expression of its values of 1 for $m=0$;  $2\cos^2 m\phi$ for $m>0$ whenever $\sigma=+$; and $2\sin^2m\phi$ for $\sigma=-$ and $m>0$. Since expression (\ref{ProbZpms}) for $P_{pm\sigma}$ depends on $l$ only through an overall coefficient and the upper integration limit, its partial derivative w.r.t.~$l$ is easy to calculate. The partial contribution of $Z_{pm\sigma}$ to the CFI matrix element $F_{ll}$ may thus be expressed as 
\ba
\label{Fll_pms}
&F_{ll}^{pm\sigma}={1\over P_{pm\sigma}}\left({\partial P_{pm\sigma}\over \partial l}\right)^2\nn
&\qquad={1\over P_{pm\sigma}}\Bigg[-{P_{pm\sigma}\over l}+{4(p+1)\over l}{J_{p+1}^2(\pi l)\over (\pi l)^2}\nn
&\times \Big\{2\delta_{\sigma,-}\sin^2 m\phi+ \delta_{\sigma,+}\left[\delta_{m0}+2\left(1-\delta_{m0}\right)\cos^2 m\phi\right]\Big\}\Bigg]^2.
\end{align}
By squaring the sum of the two terms within the large square brackets, we may expand expression (\ref{Fll_pms}) into three terms. Upon adding those three terms over all non-negative integral values of $p$, non-negative integral values of $m$ in steps of 2 starting with 0 or 1 depending on whether $p$ is even or odd and ending at $p$, and the two parity values, $\sigma=\pm$, and noting that the probabilities add up to 1, we may express the full CFI over all Zernike modes as
\ba
\label{CFIfull}
F_{ll}^{(\infty)} = &{1\over l^2}\Bigg[1-8\sum_{p=0}^\infty(p+1)^2{J_{p+1}^2(\pi l)\over (\pi l)^2}\nn
&+{4\over (\pi l)^3}\sum_{p=0}^\infty(p+1)^2{J_{p+1}^4(\pi l)\over \int_0^{\pi l} dw J_{p+1}^2(w)/w^2}\Bigg]\nn
=& {1\over l^2}\Bigg[{4\over (\pi l)^3}\sum_{p=0}^\infty(p+1)^2{J_{p+1}^4(\pi l)\over \int_0^{\pi l} dw J_{p+1}^2(w)/w^2}-1\Bigg].
\end{align}
To reach the first equality in Eq.~(\ref{CFIfull}), we used the normalization of the probability sum,
\be
\label{ProbSum}
\sum_{p,m,\sigma} P_{pm\sigma} =1,
\ee
and the fact that for fixed $p$, the sum over all allowed values of the quantity in braces in Eq.~(\ref{Fll_pms}) is simply $(p+1)$, independent of $\phi$. The final equality in Eq.~(\ref{CFIfull}) follows on using in the second term of the first equality the Gegenbauer expansion (\ref{Gegenbauer}) for $n=1$ in the special limiting case of $r\to r',\ \theta'\to 0$, namely
\be
\label{sumrule}
\sum_{p=0}^\infty(p+1)^2{J_{p+1}^2(\pi l)\over (\pi l)^2}={1\over 4},
\ee
independent of the argument $\pi l$.  

The other two elements of $\bF$ contain partial derivatives w.r.t.~$\phi$, and thus receive finite contributions only from those Zernike modes for which $m\neq 0$. All Zernike modes are separable into $l$ and $\phi$ dependent factors, with the explicitly $\phi$-dependent ones coming in unlike-parity pairs with their $\cos m\phi$ and $\sin m\phi$ angular dependences being the only difference between the two modes in each pair. The corresponding probabilities thus only differ in their angular dependences through their overall $\cos^2 m\phi$ and $\sin^2 m\phi$ factors, as in Eq.~(\ref{ProbZ1234}) for $P_2$ and $P_3$ for which $m=1$. For such separable mode pairs, since $\sin^2 m\phi=1-\cos^2m\phi$, while their $l$-dependent factors are identical, it follows that for opposite-parity modes, $Z_{pm\sigma},Z_{pm\sigma'}$, with $\sigma\neq \sigma'$,
\ba
\label{dphi_dl_P}
{\partial P_{pm\sigma}\over \partial \phi} &= -{\partial P_{pm\sigma'}\over \partial \phi};\nn
{1\over P_{pm\sigma}}{\partial P_{pm\sigma}\over \partial l} &= {1\over P_{pm\sigma'}}{\partial P_{pm\sigma'}\over \partial l}.
\end{align}
On multiplying the two identities in Eq.~(\ref{dphi_dl_P}) and transposing the two sides of the resulting identity to the same side, we see that each pair of opposite-parity modes, for fixed $p,m$, together contribute nothing to the off-diagonal element, $F_{l\phi}$, of CFI.  Adding such vanishing contributions over all allowed values of mode indices $p,m$, means that $F_{l\phi}$, just like the corresponding QFI matrix element $H_{l\phi}$, vanishes identically even when {\em all} Zernike mode projections are included.

The remaining, diagonal matrix element, $F_{\phi\phi}$, of CFI, whose inverse provides the lowest possible variance of any unbiased estimation of the orientation angle of the line source using Zernike projections, may be calculated by  noting from expression (\ref{ProbZpms}) that for $m\neq 0$,
\be
\label{dphi2_P}
\left({\partial P_{pm\pm}\over \partial \phi}\right)^2=4\sin^2 m\phi\,\cos^2 m\phi\, (P_{pm+}+P_{pm-})^2,
\ee
while 
\be
\label{Ppm}
P_{pm\pm}=(P_{pm+}+P_{pm-})\left\{
\begin{array}{l}
\cos^2m\phi\\
\sin^2 m\phi.
\end{array}
\right.
\ee
These two relations immediately yield the following expression for $F_{\phi\phi}$:
\ba
\label{Fphiphi}
F_{\phi\phi}=&\sum_{pm,\sigma} {1\over P_{pm\sigma}}\left({\partial P_{pm\sigma}\over \partial \phi}\right)^2\nn
=&4\sum_{p,m\neq 0} (P_{pm+}+P_{pm-})\nn
=&{32\over \pi l}\sum_{p=1}^\infty (p+1)\int_0^{\pi l}dw {J_{p+1}^2(w)\over w^2}\lfloor{p+1\over 2}\rfloor,
\end{align}
in which $\lfloor(p+1)/2\rfloor$, denoting the integer part of $(p+1)/2$, is the number of nonzero values of $m$ allowed for a given integer value of $p$. Since $\lfloor(p+1)/2\rfloor$ is equal to $(p+1)/2$ for odd $p$ and $p/2$ for even $p$, we may write Eq.~(\ref{Fphiphi}) as 
\ba
\label{Fphiphi2}
F_{\phi\phi}=&{16\over \pi l}\Bigg[\sum_{p=1}^\infty (p+1)^2\int_0^{\pi l}dw {J_{p+1}^2(w)\over w^2}\nn
                    &-\sum_{p=2,4,\ldots} (p+1)\int_0^{\pi l}dw {J_{p+1}^2(w)\over w^2}\Bigg]\nn
                   =&{16\over \pi l}\Bigg[\int_0^{\pi l}dw \left({1\over 4}-{J_{1}^2(w)\over w^2}\right)\nn
                    &-\sum_{p=2,4,\ldots} (p+1)\int_0^{\pi l}dw {J_{p+1}^2(w)\over w^2}\Bigg],
\end{align}
where we used identity (\ref{sumrule}), with $\pi l$ replaced by $w$, to reach the second equality from the first.

\paragraph{The Disk-Source Problem}
Use of the FT relation (\ref{ZFT}), followed by an integration over the azimuthal angle $\phi$  using the identities, 
\be
\label{cos_sin2_av} 
\int_0^{2\pi}\! d\phi\,\cos^2 \!m\phi = \int_0^{2\pi}\!d\phi\, \sin^2\! m\phi = 1/2, \ \ m\neq 0,
\ee
expresses the probability $P_j$ given by Eq.~(\ref{ZProbs1}) as
\be
\label{ZProbs}
P_{pm\sigma}(a)={2(p+1)\over \pi^2a^2}\int_0^{2\pi a} dw {J_{p+1}^2(w)\over w},
\ee
independent of $m$ and $\sigma$. In Eq.~(\ref{ZProbs}), we have, as before, switched to the full three-index notation for the Zernike modes and the corresponding probabilities, and  $w=2\pi a r$ serves as the new integration variable along the radial coordinate $r$. To evaluate the radius-estimation CFI defined as
\be
\label{Faa}
F_{aa} = \sum_{p,m,\sigma}{1\over P_{pm\sigma}}\left( {\partial P_{pm\sigma}\over \partial a}\right)^2,
\ee
we follow a procedure entirely analogous to that which led from Eq.~(\ref{Fll_pms}) to (\ref{CFIfull}) for the line-source problem considered in Sec.~III. In view of the form of expression (\ref{ZProbs}), the analog of Eq.~(\ref{Fll_pms}) is the following:
\ba
\label{Faa_pms}
F_{aa}^{pm\sigma}&={1\over P_{pm\sigma}}\left({\partial P_{pm\sigma}\over \partial a}\right)^2\nn
=&{1\over P_{pm\sigma}}\Bigg[-{2P_{pm\sigma}\over a}+{4(p+1)\over \pi a^2}{J_{p+1}^2(2\pi a)\over (2\pi a)}\Bigg]^2.
\end{align}
By squaring the terms within the large square brackets, we may expand expression (\ref{Fll_pms}) into three terms. Summing these terms over all possible integral values of $m$ in steps of 2 starting with 0 or 1 depending on whether $p$ is even or odd and ending at $p$, and the two parity values, $\sigma=\pm$, we may express the full CFI over all Zernike modes as
\ba
\label{Faa_full}
F_{aa}^{(\infty)} = &{4\over a^2}\Bigg[1-8\sum_{p=0}^\infty(p+1)^2{J_{p+1}^2(2\pi a)\over (2\pi a)^2}\nn
&+{1\over 2\pi^2 a^2}\sum_{p=0}^\infty(p+1)^2{J_{p+1}^4(2\pi a)\over \int_0^{2\pi a} dw J_{p+1}^2(w)/w}\Bigg]\nn
=& {4\over a^2}\Bigg[{1\over 2\pi^2 a^2}\sum_{p=0}^\infty(p+1)^2{J_{p+1}^4(2\pi a)\over \int_0^{2\pi a} dw \,J_{p+1}^2(w)/w}-1\Bigg].
\end{align}
In Eq.~(\ref{Faa_full}), we used the normalization (\ref{ProbSum}) of probabilities to reach the first term inside the brackets in the first equality, the fact that there are $p+1$ modes for each value of $p$ to reach the second and third term in that equality, and the sum rule (\ref{sumrule}), with $l$ replaced by $2a$, to replace the second term inside those brackets by $-2$ to reach the second equality.

\section{Bessel Form of the Generalized PSWFs and Numerical Considerations}
We now derive the Bessel form of the generalized radial PSWFs directly from the integral equation (\ref{eigenrelation2a}) upon substituting the separable form (\ref{PSWF}) of the eigenfunctions into the former. Substituting Gegenbauer's expansion (\ref{Gegenbauer}), for $n=1$ and $\theta'$ replaced by $\theta'-\theta$,
\ba
\label{BesselAddThm}
{J_1(c|\br-\brp|)\over c|\br-\brp|}=2\sum_{m=0}^\infty& (m+1){J_{m+1}(cr)\over cr}{J_{m+1}(cr')\over cr'}\nn\times& \left[{\sin(m+1)(\theta'-\theta)\over \sin(\theta'-\theta)}\right],
\end{align}
into Eq.~(\ref{eigenrelation2a}), shifting the angular integral over $\theta'$ to $\theta'+\theta$, and interchanging the order of summation and integration, we obtain the following equation:
\ba
\label{eigenrelation2}
2{\mu_N\over \pi}&\sum_{m=0}^\infty (m+1){J_1(cr)\over cr}\int_0^1 dr' r' {J_{m+1}(cr')\over cr'}\Psi_{Nn}(r') \nn
&\times \oint d\theta'{\sin(m+1)\theta'\over \sin\theta'}\left\{
\begin{array}{l}
\cos N(\theta'+\theta) \\
\sin N(\theta'+\theta)
\end{array}
\right.\nn
&=\mu_N\lambda_{Nn}\Psi_{Nn}(r)
\left\{
\begin{array}{l}
\cos N\theta\\
\sin N\theta.
\end{array}
\right.
\end{align}
That this equation holds for all values of $\theta$, in spite of the apparent disparity between its two sides, immediately follows by substituting into its left-hand side (LHS) the trigonometric relations,
\ba
\label{trig}
\cos N(\theta'+\theta) = &\cos N\theta'\cos N\theta-\sin N\theta'\sin N\theta,\nn
\sin N(\theta'+\theta) = &\sin N\theta'\cos N\theta+\cos N\theta'\sin N\theta, 
\end{align} 
and noting that the $\sin N\theta'$ terms in these relations make a vanishing contribution to the $\theta'$ integral because of the $\theta'\to -\theta'$ oddness of its integrand. In light of these observations, we immediately see that the angular dependences of the surviving integral terms are identical on both sides and thus can be suppressed, yielding the following purely radial integral equation:
\be
\label{RadialPSWF1}
\Psi_{Nn}(r)={2\over \pi\lambda_{Nn}}\sum_{m=0}^\infty c^{(Nn)}_m {J_{m+1}(cr)\over cr},
\ee
where the coefficients $c^{(Nn)}_m$ are defined by the integral relation
\ba
\label{coeff1}
c^{(Nn)}_m = (m+1)&\int_0^1 dr' r' {J_{m+1}(cr')\over cr'}\Psi_{Nn}(r')\nn
\times&\oint d\theta'{\sin(m+1)\theta'\over \sin\theta'}\cos N\theta'.
\end{align}
Note that the coefficients $c^{(Nn)}_m$ vanish unless $m-N$ is even, since under the shift $\theta'\to\theta'+\pi$, the integrand of the $\theta'$ integral is odd when $m-N$ is odd.  A second trigionometric identity, which follows simply by using the geometric-series sum formula applied to terms that are powers of $\exp(2i\theta)$, namely
\be
\label{trig2}
{\sin (m+1)\theta\over \sin\theta}= \left\{\displaystyle{
\begin{array}{ll}
1+2\sum_{k=1}^{m/2} \cos 2k\theta, & m:\ {\rm even}\\
2\sum_{k=0}^{(m-1)/2} \cos (2k+1)\theta, &m:\ {\rm odd},
\end{array}}
\right.
\ee
when used in conjunction with the Fourier-series orthogonality formula,
\be
\label{FourierOrthogonality}
\oint \cos k\theta\, \cos l\theta\, d\theta=(1+\delta_{k0})\pi \delta_{kl},
\ee
shows immediately that $c^{(Nn)}_m$ given by Eq.~(\ref{coeff1}) must vanish unless $m\geq N$. In view of these two properties of $c^{(Nn)}_m$, we may write expression (\ref{RadialPSWF1}) as
\be
\label{RadialPSWFBesselForm2}
\Psi_{Nn}(r)=\sum_{k=0}^\infty d^{(Nn)}_k {J_{N+2k+1}(cr)\over cr},
\ee
where $d^{(Nn)}_k=2c^{(Nn)}_{N+2k}/(\pi\lambda_{Nn})$ relabels the coefficients in a simpler notation. 

A substitution of form (\ref{RadialPSWFBesselForm2}) for the radial PSWF in Eq.~(\ref{coeff1}) yields the following infinite system of equations for the coefficients $d^{(Nn)}_k$:
\be
\label{CoeffSystem}
{\pi\lambda_{Nn}\over 4}d^{(Nn)}_k = \sum_{k'=0}^\infty M^{(N)}_{kk'}d_{k'}^{(Nn)},\ k=0,1,\ldots,\infty,
\ee
where the matrix elements are defined as
\be
\label{MatEl}
M^{(N)}_{kk'}\!\!=\!\! (N+2k+1)\,A_{Nk} \int_0^1\!\! dr' r' {J_{N+2k+1}(cr')J_{N+2k'+1}(cr')\over (cr')^2},
\ee
in which the quantities $A_{Nk}$ denote the angular integrals,
\be
\label{AngInt}
A_{Nk}=\oint d\theta{\sin(N+2k+1)\theta\over \sin\theta}\cos N\theta.
\ee
Use of the trigonometric sum identity (\ref{trig2}), followed by a use of the Fourier orthogonality relation (\ref{FourierOrthogonality}), inside the integral (\ref{AngInt}) evaluates it as the constant $2\pi$,
\be
\label{AngInt2}
A_{Nk}=2\pi,
\ee 
for all non-negative integer values of $N$ and $k$. 

The matrix with elements (\ref{MatEl}) constitutes an infinite set of linearly coupled equations with a non-symmetric system matrix ${\bf M}^{(N)}$. By rescaling the coefficients $d_k^{(Nn)}$ as
\be
\label{d2f}
f_k^{(Nn)}={d_k^{(Nn)}\over \sqrt{N+2k+1}},
\ee
we may, however, transform the system (\ref{CoeffSystem}) to the form,
\be
\label{CoeffSystem2}
\lambda_{Nn}f^{(Nn)}_k = \sum_{k'=0}^\infty \tilde M^{(N)}_{kk'}f_{k'}^{(Nn)},\ k=0,1,\ldots,\infty,
\ee 
that involves a symmetric, positive-semidefinite matrix $\tilde{\bf M}^{(N)}$ with elements
\ba
\label{MatEl2}
\tilde M^{(N)}_{kk'}= &8\sqrt{(N+2k+1)(N+2k'+1)}\nn
\times & \int_0^1 dr' r' {J_{N+2k+1}(cr')J_{N+2k'+1}(cr')\over (cr')^2}
\end{align}
and guaranteed non-negative eigenvalues.

Note that since $0\leq r,r'\leq 1$, the Bessel functions inside the integral (\ref{MatEl}) decay with $k,k'$ rapidly when they exceed a number of order $(c-N-1)/2$. From Eq.~(\ref{CoeffSystem}), it then follows that the coefficients $d_k^{(Nn)}$ are also small for such values of $k$, allowing one to calculate the eigenvalues $\lambda_{Nn}$ by truncating the linear system (\ref{CoeffSystem}) of equations for the coefficients at a relatively small order, and then requiring that the underlying system matrix have a vanishing determinant. In our numerical evaluations of QFI, since we needed to achieve a very high precision, we typically truncated the infinite matrix to a $K\!\times\! K$ square matrix by allowing $k,k'$ to run from 0 to $K-1$ with $K$ of order 1000, for which the first 10-20 most significant eigenvalues are determined accurately to about ten decimal places for each value of $N$.

It is worth noting that the same approach of expanding the PSWFs in Bessel functions remains useful for 1D and higher-dimensional spaces as well. In particular, the 1D PSWFs emerge via the Gegenbauer expansion of $J_{1/2}(c|x-x'|)/|x-x'|^{1/2}$, which is proportional to the kernel, $\sin (c|x-x'|)/(\pi |x-x'|)$, of the integral operator of which they are the eigenfunctions over the interval $(-1,1)$,
in terms of Bessel functions of half integer order. In the Bessel form, the expansion of the PSWFs holds its validity for values of the argument $x$ outside the interval $(-1,1)$ over which these functions are typically defined. This would have dispensed with the 1D discrete PSWF sequences, which are only defined over the interval $(-1,1)$, that we utilized in Ref.~\cite{Prasad20b} for calculating QFI for super-localizing and super-resolving a pair of incoherent sources in two dimensions as a function of the source emission bandwidth.  

\paragraph{Closed-Form Evaluation of Matrix Elements $M^{(N)}_{kk'}$} Our numerical computations of the eigenvalues and associated eigenvectors of $\tilde {\bf M}^{(N)}$ become highly efficient when we recognize that we may analytically evaluate the integral in expression (\ref{MatEl2}) for each matrix element. We start with the Bessel differential equation for two different orders, say $\mu$ and $\nu$,
\ba
\label{BesselEq}
xJ_\mu^{''}+J'_\mu+(x-\mu^2/x)J_\mu(x)=&0;\nn
xJ_\nu^{''}+J'_\nu+(x-\nu^2/x)J_\mu(x)=&0;
\end{align}
in which each prime indicates a single derivative with respect to the argument $x$ of each Bessel function,
and then take their difference after multiplying the first by $J_\nu$ and the second by $J_\mu$. This difference may be expressed as 
\be
\label{BesselEqDiff}
\left[x\left( J'_\mu J_\nu-J'_\nu J_\mu\right)\right]'=(\mu^2-\nu^2){J_\mu J_\nu\over x}.
\ee
Integrating both sides of this equation w.r.t. $x$ from 0 to $c$ permits an evaluation of the integral of its RHS, since its LHS is a total derivative and so easily integrated as
\be
\label{BesselProdInt}
\int_0^c {J_\mu(x) J_\nu(x)\over x}dx = c{\left[ J'_\mu (c) J_\nu (c) - J'_\nu(c) J_\mu(c)\right]\over \mu^2-\nu^2},\ \mu\neq \nu.
\ee
This expression can be simplified by using the Bessel identity \cite{GR96},
\be
\label{BesselId}
J'_\mu=-\mu{J_\mu\over x}+J_{\mu-1},
\ee to its final form that does not involve any derivatives and is easily evaluated numerically,
\ba
\label{BesselProdInt1}
\int_0^c &{J_\mu(x) J_\nu(x)\over x}dx = -{J_\mu(c)J_\nu(c)\over \mu+\nu}\nn
&+c{\left[ J_{\mu-1} (c) J_\nu (c) - J_{\nu-1}(c) J_\mu(c)\right]\over \mu^2-\nu^2},\ \mu\neq \nu.
\end{align}
Note that all the off-diagonal matrix elements $M^{(Nn)}_{kk'}$ given by Eq.~(\ref{MatEl}), $k\neq k'$, corresponding to $\mu\neq \nu$ in expression (\ref{BesselProdInt1}), decrease super-exponentially with $k,k'$ whenever $k,k'>>(c-N-1)/2$, since all Bessel functions occurring in that relation become super-exponentially small in this limit.
 
For $\mu=\nu=n\neq 0$, we may evaluate the integral by using l'Hospital rule, but that requires taking derivative of Bessel functions w.r.t. their order. A simpler approach makes use of identity (\ref{BesselId}) to reach the integral identity,
\be
\label{BesselInt}
\int_0^c {J_n^2(x)\over x} dx= {1\over n}\int_0^c \left[-J_n(x)J'_n(x)+J_n(x)J_{n-1}(x)\right] dx,
\ee
in which the first term on the RHS is the integral of the total derivative $(1/2)[J_n^2(x)]'$, which is easily evaluated, so we have
\be
\label{BesselInt2}
\int_0^c {J_n^2(x)\over x} dx= -{1\over 2n}J_n^2(c)+{1\over n}\int_0^c J_n(x)J_{n-1}(x)\, dx,
\ee
in which the integral on the RHS still needs to be evaluated. But this requires no integration when we note that it may be expressed in terms of an integral with $n\to n-1$, a procedure that can be iterated down to $n=1$ when the starting value of $n$ is a positive integer. We illustrate this iterative procedure by using the notation,
\be
\label{BesselInt3}
G_n\defeq \int_0^c J_n(x)\, J_{n-1}(x)\, dx,
\ee
and using the Bessel identity \cite{GR96}, $J_n=-2J'_{n-1}+J_{n-2}$, followed by an integration of a total derivative to derive the recursion relation,
\be
\label{recursion}
G_n =G_{n-1}-J^2_{n-1}(c)+\delta_{n1},
\ee
the Kronecker $\delta$ term resulting from the fact that $J_0(0)=1$. Use of this recursion relation iteratively evaluates $G_n$ as the sum
\be
\label{Gn}
G_n=G_1-\sum_{k=1}^{n-1}J_k^2(c)={1\over 2}[1-J_0^2(c)]- \sum_{k=1}^{n-1}J_k^2(c),
\ee
where we used the fact that $J_1=-J'_0$ to evaluate $G_1$ as the integral of the total derivative of $-(1/2)J^2$ and thus equal to $[1-J_0^2(c)]/2$. Use of this identity then evaluates the integral in Eq.~(\ref{BesselInt2}) as the sum
\be
\label{BesselInt4}
\int_0^c {J_n^2(x)\over x} dx= {1\over 2 n}\left[1-J_0^2(c)- J_n^2(c)\right]-{1\over n}\sum_{k=1}^{n-1}J_k^2(c).
\ee
In view of the summation formula \cite{GR96},
\be
\label{BesselSumFormula}
1-J_0^2(c)=2\sum_{k=1}^\infty J_k^2(c),
\ee
it then follows from Eq.~(\ref{BesselInt4}) that
\be
\label{BesselInt5}
\int_0^c {J_n^2(x)\over x} dx= {1\over 2n}J_n^2(c)+{1\over n}\sum_{k=n+1}^\infty J_k^2(c),\ n\geq 1,
\ee
which also decays super-exponentially with the index $n$ in the limit of large $n$, specifically when $n>>c$ for which all the Bessel functions occurring in this relation decrease super-exponentially. 

\paragraph{Orthonormality of Eigenfunctions} In view of relation (\ref{d2f}) and expansion (\ref{RadialPSWFBesselForm2}) for the radial eigenfunctions, the LHS of the orthonormality condition (\ref{RadialEigenNorm}) reduces to the form
\ba
\label{RadialEigenNormLHS}
\int_0^1 dr\, r\, \Psi_{Nn}(r)\,\Psi_{Nn'}(r)=&{1\over 8} \sum_{k,k'} f_k^{(Nn)}f_{k'}^{(Nn')} \tilde M_{kk'}^{(N)}\nn
                                                                 =&{\lambda_{Nn'}\over 8}\sum_k f_k^{(Nn)} f_k^{(Nn')}\nn
                                                                 =&{\lambda_{Nn'}\over 8}\underline{f}^{(Nn)T}\underline{f}^{(Nn')},
\end{align}
in which we used expression (\ref{MatEl2}) to arrive at the double sum of the first line and eigen-relation (\ref{CoeffSystem2}) to arrive at the second line, which may be expressed as the inner product of the column vectors of coefficients of the two different eigenvectors, as in the third line in which the superscript $T$ on the first vector denotes its matrix transposition. In view of the RHS of the orthonormality condition (\ref{RadialEigenNorm}), we see from expression (\ref{RadialEigenNormLHS}) that the coefficient vectors corresponding to two different eigenfunctions must obey the matrix-product orthogonality requirement,
\be
\label{RadialCoeffNorm}
 \underline{f}^{(Nn)T}\underline{f}^{(Nn')}={8\pi\over \lambda_{Nn}^2}\delta_{nn'},\  \forall N.
\ee
Since the typical matrix eigensolver, such as {\it eig} in Matlab, normalizes the coefficient vectors to have unit norm, we must first scale the so-normalized coeffcient eigenvector by the factor $(8\pi)^{1/2}/\lambda_{Nn}$ before using the relation (\ref{d2f}) to arrive at the vector of coefficients $d_k^{(Nn)}$ and then substituting the latter into expansion (\ref{RadialPSWFBesselForm2}) to arrive at the final form of what is the unit-norm radial eigenfunction $\Psi^{(Nn)}(r)$.

\section{Evaluation of $L_{Nn}^{(p)}$, $P_a$, and $Q_a$ Functions}
We first evaluate $P_a$ given by the second of the expressions (\ref{LPQ}) by employing the pupil-plane wavefunction formula (\ref{wavefunction}) as the pupil-plane integral,
\ba
\label{Pintegral}
P_a(\brp,&\br)= -2i \int_0^1 du\, u \oint d\theta_u \bu\cdot\br\, \exp[-i2\pi a\bu\cdot(\br-\brp)]\nn
  =& -2i r\int_0^1 du\, u^2 \oint d\theta_u \cos(\theta_u-\theta)\nn
    &\qquad\qquad\times \exp[-i2\pi a u|\br-\brp|\cos(\theta_u-\theta_{rr'})]\nn
    =& -4\pi r\cos(\theta-\theta_{rr'}) \int_0^1 du\, u^2 J_1(2\pi a u|\br-\brp|)\nn
    =&-4\pi {\br\cdot(\br-\brp)\over |\br-\brp|} {J_2(2\pi a|\br-\brp|)\over 2\pi a|\br-\brp|}\nn
    =&-8\pi^2 a [r^2-rr'\cos(\theta'-\theta)] {J_2(2\pi a|\br-\brp|)\over (2\pi a|\br-\brp|)^2},
\end{align}
in which the symbol $\theta_{rr'}$ denotes the polar angle of the vector $(\br-\brp)$. To arrive at the third and fourth relations in Eq.~(\ref{Pintegral}), we used successively the following two Bessel integral identities:
\ba
\label{BesselIntId12}
{i^m\over 2\pi}\oint& d\phi \cos m\phi \exp[-iz\cos(\phi+\delta\phi)]= \cos m\delta\phi \,J_m(z),\nn
\int& dz\, z^n J_{n-1}(z)=z^n J_n(z),
\end{align}
with $m=1,\phi=\theta_u-\theta, \delta\phi=\theta-\theta_{rr'}, n=2$, and then used the fact that the inner product of vectors $\br$ and $\br-\brp$ is simply the product of their magnitudes and cosine of the angle between the two, the latter being $\theta-\theta_{rr'}$. The first identity in Eq.~(\ref{BesselIntId12}) is obtained by noting that the exponential inside the integrand is the generating function for Bessel functions in powers of $\exp[i(\phi+\delta \phi)]$, while the second identity is the integral form of the relation \cite{GR96},
\be
\label{BesselId1}
{d\over dz}[z^n J_{n}(z)]=z^n J_{n-1}(z).
\ee
The final relation of Eq.~(\ref{Pintegral}) follows from performing the inner product in its previous relation explictly. 
Note that since the angular dependence of $|\br-\brp|$ is also a function of $\cos(\theta'-\theta)$ alone, the final expression for $P_a$ in Eq.~(\ref{Pintegral}) depends on angles only through $\cos(\theta'-\theta)$.  

To evaluate $Q_a$ defined in Eq.~(\ref{LPQ}), we start with the wavefunction (\ref{wavefunction}) using which we may write for $Q_a$,
\ba
\label{Qintegral}
Q_a&(\brp,\br)= 4\pi \int_0^1 \!\!\!du\, u \!\oint \!d\theta_u (\bu\cdot\br)\, (\bu\cdot\brp)\nn
&\qquad\qquad\times \exp[-i2\pi a\bu\cdot(\br-\brp)]\nn
  =& 4\pi rr'\int_0^1 du\, u^3 \oint d\theta_u \cos(\theta_u-\theta)\cos(\theta_u-\theta')  \nn
    &\times\exp[-i2\pi a u|\br-\brp|\cos(\theta_u-\theta_{rr'})]\nn
    =& 2\pi rr' \int_0^1 du\, u^3 \oint d\theta_u [\cos(2\theta_u-\theta-\theta')+\cos(\theta-\theta')]\nn
    &\qquad \qquad\qquad\times\exp[-i2\pi a u|\br-\brp|\cos(\theta_u-\theta_{rr'})]\nn
    =&4\pi^2 rr'\Bigg[\!\!-\cos(2\theta_{rr'}-\theta-\theta')\int_0^1 \!\!du\, u^3 J_2(2\pi a|\br-\brp|u)\nn
      &\qquad\quad+\cos(\theta-\theta')\int_0^1 du\, u^3 J_0(2\pi a|\br-\brp|u)\Bigg]\nn
    =&4\pi^2 rr'\Bigg[-\cos(2\theta_{rr'}-\theta-\theta'){J_3(X)\over X}\nn
      &\qquad\quad+\cos(\theta-\theta')\left({J_1(X)\over X}-2{J_2(X)\over X^2}\right)\Bigg],
\end{align}
where $X$ was defined earlier in Eq.~(\ref{Pintegral1}). In Eq.~(\ref{Qintegral}), the second equality follows from the first when we use the definition of the inner product of two vectors as the product of their magnitudes and cosine of the angle between them. The third equality results on applying the trigonometric identity, $2\cos A\, \cos B=\cos(A+B)+\cos(A-B)$, in the second equality, while the fourth follows from the third when using the first of the Bessel identities (\ref{BesselIntId12}) for $m=2$ and $2\phi=2\theta_u-(\theta+\theta'), 2\delta\phi=(\theta+\theta'-2\theta_{rr'})$. The first term in the final equality in Eq.~(\ref{Qintegral}) follows from the use of the second of the Bessel integral identities (\ref{BesselIntId12}) for $n=3$, while the second term is a result of another Bessel identity,
\ba
\label{BesselId1}
\int dz\, z^3 J_0(z)= &\int dz\, z^2 {d\over dz}(z J_1)\nn
                             =&-z^3 J_1-2\int dz\, z^2 J_1(z)\nn
                             =&-z^3J_1-2 z^2 J_2,
\end{align}
in which applying the second identity in (\ref{BesselIntId12}) twice yields the first and third equalities, with the second resulting from an integration by parts.

The first cosine term in Eq.~(\ref{Qintegral}) may be written, using a trigonometric sum formula, as 
\ba
\label{Qcosterm1}
\cos&(2\theta_{rr'}-\theta-\theta')=\cos(\theta-\theta_{rr'})\cos(\theta'-\theta_{rr'})\nn
&\qquad\qquad -\sin(\theta-\theta_{rr'})\sin(\theta'-\theta_{rr'})\nn
=&{\br\cdot(\br-\brp)\over r|\br-\brp|}{\brp\cdot(\br-\brp)\over r'|br-\brp|}-{|\br\times(\br-\brp)|\over r|br-\brp|}{|\brp\times(\br-\brp)|\over r'|\br-\brp|}\nn
=&{[r-r'\cos(\theta-\theta')][r\cos(\theta-\theta')-r']-rr'\sin^2(\theta-\theta')\over |\br-\brp|^2}\nn
=&{(r^2+r^{'2})\cos(\theta'-\theta)-2rr'\over r^2+r^{'2}-2rr'\cos(\theta'-\theta)}\nn
=&c^2{(r^2+r^{'2})\cos(\theta'-\theta)-2rr'\over X^2},
\end{align}
in which we used simple vector identities, $\bA\cdot\bB =AB\cos\theta_{AB}$, $|\bA\times\bB|=AB\sin \theta_{AB}$, $\bA\cdot\bA=A^2$, and $\bA\times\bA=0$, for any two vectors $\bA,\bB$ with angle $\theta_{AB}$ between them and the definition of symbol $X$ defined in Eq.~(\ref{Pintegral1}). In view of Eq.~(\ref{Qcosterm1}), we may express $Q_a$ of Eq.~(\ref{Qintegral}) finally as
\ba
\label{Qintegral1}
Q_a&(\brp,\br)= 4\pi^2 rr'\Bigg\{c^2[2rr'-(r^2+r^{'2})\cos(\theta'-\theta)]{J_3(X)\over X^3}\nn
                       &\qquad+\cos(\theta'-\theta)\left({J_1(X)\over X}-2{J_2(X)\over X^2}\right)\Bigg\}.
\end{align}

If we now substitute for $\la \lambda_{Nn}|$ the Hermitian adjoint of Eq.~(\ref{expansion}) into the first of the expressions (\ref{LPQ}) and use the second of them, then we find 
\be
\label{LNn}
L_{Nn}^{(p)}(\br) = {1\over\pi}\int_{\cD_1}dA' C_{Nn}^{(p)}(r',\theta')\, P_a(\brp,\br).
\ee
The angular integral over $\theta'$ in this expression can be performed by shifting $\theta'\to\theta'+\theta$ inside its integrand. This means either a $\cos N(\theta'+\theta)$ or $\sin N(\theta+\theta')$ angular dependence for the coefficient function $C_{Nn}^{(\pm)}(r,\theta'+\theta)$ according to its expression (\ref{PSWF}), while the rest of the integrand depends only on $\cos\theta'$, following such an angular shift. Since  $\cos\theta'$ is even while the second terms on the RHS of the identities below, namely
\ba
\label{CS}
\cos N(\theta+\theta') =&\cos N\theta\, \cos N\theta' - \sin N\theta\, \sin N\theta',\nn 
\sin N(\theta+\theta') =&\sin N\theta\, \cos N\theta' + \cos N\theta\, \sin N\theta',
\end{align}
are odd under reflection in the $x$ axis, $\theta'\to-\theta'$, only the first terms on their RHSs, which comprise the net angular dependences of $C_{Nn}^{(\pm)}(\theta+\theta')$, can contribute to the angular part of the integral (\ref{LNn}). In other words, $L_{Nn}^{(p)}(\br)$ also has the same angular dependences, namely $\cos N\theta$ or $\sin N\theta$, corresponding to the eigenfunctions $C_{Nn}^{(p)}(\brp)$ inside integral (\ref{LNn}). In view of this fact, the integrand on the RHS of the first disk integral in Eq.~(\ref{prhoMatEl}) will vanish unless $N=N'$ and the two states $|\lambda_{Nn}\ra$ and $|\lambda_{N'n'}\ra$ also have the same parity under reflection in the $x$ axis, $p=p'$, since unequal-parity trigonometric functions are always orthogonal over the $2\pi$ angular period,
\be
\label{CSorthogonality}
\oint d\theta \cos N\theta\, \sin N'\theta = 0, \ \forall \ {\rm integral}\ N,N'.
\ee
This proves our assertion, and the first of the matrix elements in Eq.~(\ref{prhoMatEl}), after the angular integral is trivially performed, may be written in terms of a simple radial integral over the unit disk as
\ba
\label{prhoMatElB1}
\la \lambda_{Nn}^{(p)}|\partial\hrho_2&|\lambda_{N'n'}^{(p')}\ra =\delta_{NN'}\delta_{pp'} {1\over \pi}\int_0^1 dr\, r\,\big [ \lambda_{Nn'}\Psi_{Nn'}(r)\nn
&\times L_{Nn}(r)+\lambda_{Nn}\Psi_{Nn}(r)L_{Nn'}(r) \big],
\end{align}
in which $\Psi_{Nn}(r)$ and $L_{Nn}(r)$ are simply the radial parts of the full functions, $C_{Nn}^{(p)}(\br)$ and $L_{Nn}^{(p)}(\br)$, respectively. The first of these radial functions is defined in Eq.~(\ref{PSWF}), while the second is the radial part of expression (\ref{LNn}), {\em i.e.,}
\be
\label{rLNn}
L_{Nn}(r) = {1\over\pi}\int_0^1 dr'\, r' \Psi_{Nn}(r')\oint d\theta' P_a(r',r,\theta')) \cos N\theta',
\ee
in which the quantity $P_a(r',r,\theta')$ is simply the value of $P_a(\brp,\br)$ evaluated at $\theta=0$, which from Eq.~(\ref{Pintegral}) takes the following form, when expressed in terms of $c=2\pi a$:
\ba
\label{Pintegral1}
&P_a(r',r,\theta')=-4\pi c (r^2-rr'\cos\theta') {J_2(X_0)\over X_0^2}, \nn
&\ \ X\defeq c\sqrt{r^{'2}+r^2-2rr'\cos(\theta'-\theta)},\ \ X_0=X\vert_{\theta=0}.
\end{align}


\begin{thebibliography}{99}
\bibitem{footnote1} For thresholds below $8\times10^{-10}$, we could not evaluate the eigenvalues that were still smaller sufficiently accurately, but the general trends held as we lowered the threshold from larger values down to the chosen threshold of $10^{-9}$.
\end{thebibliography}
\end{document}